\documentclass[preprint,10pt]{elsarticle}

\usepackage{graphicx}
\usepackage{amssymb}
\usepackage{amsthm}
\usepackage{amsmath}

\DeclareMathOperator{\tr}{tr}
\usepackage{bm}

\usepackage{xcolor}
\usepackage[margin=2.5cm]{geometry}
\usepackage{mathtools}  
\usepackage{xfrac}  
\usepackage{upgreek}
\usepackage{soul}
\usepackage{cancel}
\usepackage{accents}
\usepackage{bigints}
\usepackage{subcaption}
\usepackage{tikz}
\usepackage{verbatim}
\usepackage{todonotes}
\usepackage{mathrsfs}

\newtheorem{thm}{Theorem}[section]
\theoremstyle{definition}
\newtheorem{remark}[thm]{Remark}

\usepackage[%
    colorlinks=false,%
    hyperindex,%
    plainpages=false,%
    bookmarksopen,%
    bookmarksnumbered]{hyperref}
\biboptions{numbers,sort&compress}

\usepackage{lineno}  
\usepackage{stmaryrd}
\usepackage[framed]{matlab-prettifier}
\usepackage[normalem]{ulem}
\usepackage{overpic}

\newcommand{\beq}{\begin{equation}}
\newcommand{\eeq}{\end{equation}}

\newcommand{\dt}{\partial_t}

\def\rspace{\mathbb{R}}
\def\map{\bm{\phi}}

\def\spatdom{\Omega_t}
\def\refdom{\Omega_0}
\def\spatdomf{\Omega_t^f}

\def\refdoms{\Omega_0^s}

\def\spatdomu{\hat{\Omega}_t}
\def\refdomu{\hat{\Omega}_0}
\def\spatdomfu{\hat{\Omega}_t^f}
\def\spatdomsu{\hat{\Omega}_t^s}
\def\refdomfu{\hat{\Omega}_0^f}
\def\refdomsu{\hat{\Omega}_0^s}
\def\sfintu{\hat{\Gamma}_t^{sf}}
\def\fintu{\hat{\Gamma}_t^{f}}
\def\sfintrefu{\hat{\Gamma}_0^{sf}}
\def\sintrefu{\hat{\Gamma}_0^{s}}

\def\fint{\Gamma_t^{f}}

\def\sintref{\Gamma_0^{s}}
\def\sfint{\Gamma_t^{sf}}
\def\sfintref{\Gamma_0^{sf}}

\def\fintdir{\Gamma_{t, D}^{f}}
\def\fintneu{\Gamma_{t, N}^{f}}
\def\fintaxi{\Gamma_{t, A}^{f}}
\def\sintdir{\Gamma_{0, D}^{s}}
\def\sintneu{\Gamma_{0, N}^{s}}

\def\normf{\bm{n}^f}

\def\normsf{\bm{n}^{sf}}
\def\normsref{\bm{n}_0^s}

\def\lbr{\left(}
\def\rbr{\right)}

\newcommand{\eps}{\hat{\epsilon}}

\def\glenergy{\hat{\Psi}^f}
\def\velf{\bm{v}}
\def\phase{c}
\def\viscf{\eta}
\def\viscfa{\hat{\eta}_\mathrm{a}}
\def\viscfl{\hat{\eta}_\mathrm{l}}
\def\densf{\rho}
\def\densfa{\hat{\rho}_\mathrm{a}}
\def\densfl{\hat{\rho}_\mathrm{l}}

\def\fluidsf{\hat{\gamma}_\mathrm{la}}
\def\sigmaf{\bm{\sigma}^f}

\def\dw{D_w}

\def\disps{\bm{u}}

\def\vels{\dt{\disps}}
\def\denss{\hat{\rho}_0^s}

\def\sigmas{\bm{\sigma}^s}
\def\pkone{\bm{P}}
\def\defgrad{\bm{F}}

\def\xm{{\bm{X}}}

\def\capno{\mathrm{Ca}}
\def\capnotip{\mathrm{Ca_t}}
\def\capnocl{\mathrm{Ca_{cl}}}
\def\capnocr{\mathrm{Ca_{cr}}}
\def\capnoavg{\overline{\mathrm{Ca}}}

\def\ohno{\text{Oh}}
\def\pecno{\text{Pe}}
\def\cahnno{\text{Cn}}

\def\xr{r}
\def\xz{z}
\def\Xr{R}
\def\Xz{Z}
\def\xrd{\hat{r}}
\def\xzd{\hat{z}}
\def\Xrd{\hat{R}}
\def\Xzd{\hat{Z}}
\def\vr{v_r}
\def\vz{v_z}
\def\vrm{\widetilde{v}_r}
\def\vzm{\widetilde{v}_z}
\def\ur{u_R}
\def\uz{u_Z}
\def\sigmafrr{\sigma^f_{\xr \xr}}
\def\sigmafrz{\sigma^f_{\xr \xz}}
\def\sigmafzz{\sigma^f_{\xz \xz}}
\def\sigmaftt{\sigma^f_{\vartheta \vartheta}}
\def\pkonerr{P_{\Xr \Xr}}
\def\pkonerz{P_{\Xr \Xz}}
\def\pkonezz{P_{\Xz \Xz}}
\def\pkonett{P_{\vartheta \vartheta}}

\def\wp{w_p}
\def\wph{w_p^h}
\def\wpd{w_p^{\prime}}

\def\wvr{w_{vr}}
\def\wvrh{w_{vr}^h}
\def\wvrd{w_{vr}^{\prime}} 

\def\wvz{w_{vz}}
\def\wvzh{w_{vz}^h}
\def\wvzd{w_{vz}^{\prime}} 

\def\wc{w_{\phase}}
\def\wch{w_{\phase}^h}
\def\wcd{w_{\phase}^{\prime}} 

\def\wmu{w_{\mu}}
\def\wmuh{w_{\mu}^h}
\def\wmud{w_{\mu}^{\prime}} 

\def\wur{w_{ur}}

\def\wuz{w_{uz}}

\def\xf{\bm{\mathcal{X}}^f}
\def\xfh{\bm{\mathcal{X}}^{f,h}}
\def\xfd{\bm{\mathcal{X}}^{f,'}}

\def\yf{\bm{\mathcal{Y}}^f}
\def\yfh{\bm{\mathcal{Y}}^{f,h}}
\def\yfd{\bm{\mathcal{Y}}^{f,'}}

\def\ph{p^h}
\def\vrh{\vr^h}
\def\vzh{\vz^h}

\def\phaseh{\phase^h}
\def\muh{\mu^h}
\def\vrmh{\tilde{v}_r^h}
\def\vzmh{\tilde{v}_z^h}

\def\taum{\tau_{m}}
\def\tauc{\tau_c}

\def\arclen{L_\text{arc}}
\def\axialcli{L_\text{cli}}
\def\appcont{\theta_\text{a}}

\newcommand{\quotes}[1]{``$#1$''}

\def\@fnsymbol#1{\ensuremath{\ifcase#1\or *\or \dagger\or \ddagger\or
   \mathsection\or \mathparagraph\or \|\or **\or \dagger\dagger
   \or \ddagger\ddagger \else\@ctrerr\fi}}
\newcommand{\ssymbol}[2]{^{\@fnsymbol{#2}}}

\newcommand{\sbr}{\textcolor{black}}

\let\oldbibliography\bibliography
\renewcommand{\bibliography}[1]{{\tiny\oldbibliography{#1}}}
\journal{Journal}

\begin{document}

\begin{frontmatter}




\title{Simulating fluid-fluid displacement in a soft capillary tube: How compliance delays interfacial instability and bubble pinch-off}

\author[inst1]{Sthavishtha R. Bhopalam}
\author[inst2,inst3]{Ruben Juanes}
\author[inst1]{Hector Gomez\corref{cor1}}
\address[inst1]{School of Mechanical Engineering, Purdue University, West Lafayette, IN 47907, USA.}
\address[inst2]{Department of Civil and Environmental Engineering, Massachusetts Institute of Technology, Cambridge, MA 02139, USA.}
\address[inst3]{Department of Earth, Atmospheric and Planetary Sciences, Massachusetts Institute of Technology, Cambridge, MA 02139, USA.}
\cortext[cor1]{Corresponding author email: \href{hectorgomez@purdue.edu}{hectorgomez@purdue.edu}}

\begin{abstract}

The displacement of a more viscous fluid by a less viscous immiscible fluid in confined geometries is a fundamental problem in multiphase flows. Recent experiments have shown that such fluid-fluid displacement in micro-capillary tubes can lead to interfacial instabilities and, eventually bubble pinch-off. A critical yet often overlooked aspect of this system is the effect of tube’s deformability on the onset of interfacial instability and bubble pinch-off. Here, we present a computational fluid-structure interaction model and an algorithm to simulate this fluid-fluid displacement problem in a soft capillary tube. We use a phase-field model for the fluids and a nonlinear hyperelastic model for the solid. Our fluid-structure interaction formulation uses a boundary-fitted approach and we use Isogeometric Analysis for the spatial discretization. Using this computational framework, we study the effects of inlet capillary number and the tube stiffness on the control of interfacial instabilities in a soft capillary tube for both imbibition and drainage. We find that the tube compliance delays or even suppresses the interfacial instability and bubble pinch-off---a finding that has important implications for flow in soft porous media, bio-microfluidics, and manufacturing processes.

\end{abstract}


\end{frontmatter}


\section{Introduction}
\label{sec:intro}

The displacement of a fluid by another immiscible fluid in a confined geometry, such as a capillary tube or a porous medium, is important in many applications, including enhanced oil recovery \cite{orr_taber_sci_1984}, inkjet printing \cite{yan_etal_aplm_2020}, and microfluidics \cite{whitesides_nat_2006}. This problem has been extensively studied through experiments \cite{hoffman_jcis_1975, bretherton_jfm_1961, taylor_jfm_1961}, theoretical analyses \cite{huh_jfm_1977}, and numerical simulations \cite{zhou_prl_1990, ruiz_etal_jfm_2022}. While the displacement of a less viscous fluid by a more viscous fluid (a {\it favourable} displacement problem) has been well studied, the reverse process---displacing a more viscous fluid with a less viscous one (an {\it unfavourable} displacement problem) has received less attention. This latter problem was initially explored by Taylor \cite{taylor_jfm_1961} and Bretherton \cite{bretherton_jfm_1961} in the complete wetting regime. Only more recently has the unfavourable fluid–fluid displacement problem been investigated in the partial wetting regime \cite{zhao_etal_prl_2018, pahlavan_etal_pnas_2019, esmaeilzadeh_etal_pre_2020, gao_etal_jfm_2019, suo_wrr_2024}, where it has been shown that the fluid–fluid interface remains stable at low injection flow rates, but becomes unstable---eventually leading to bubble pinch-off---when the flow rate exceeds a critical value.  The onset of interfacial instability and subsequent bubble pinch-off in a capillary tube is primarily governed by the competition between viscous forces and capillary forces. These forces can be modified, for example, by varying the flow rate of the injected fluid \cite{zhao_etal_prl_2018}, the tube’s wettability \cite{esmaeilzadeh_etal_pre_2020, gao_etal_jfm_2019}, and the tube’s geometry \cite{suo_wrr_2024}. Importantly, this form of interfacial instability differs fundamentally from the classical viscous fingering phenomenon observed in Hele-Shaw cells \cite{saffman_prs_1958}.

Several computational methods have been developed to model fluid–fluid displacement in capillary tubes. One possible classification of these methods is into sharp-interface \cite{esmaeilzadeh_etal_pre_2020, spelt_jcp_2005, sui_arfm_2014, zhang_jcp_2020} and diffuse-interface (or phase-field) models \cite{jacqmin_jcp_1999, jacqmin_jfm_2000, yue_etal_jfm_2010, yue_feng_pof_2011, gao_etal_jfm_2019, prokopev_pre_2019}. Among these methods, the phase-field approach \cite{anderson_arfm_1998, gomez_zee_2017} has become very relevant for several reasons. First, phase-field methods avoid contact-line stress singularities. Second, they naturally handle topological changes such as interfacial breakup and coalescence. Third, they can describe the dynamic wetting behavior, moving contact lines \cite{jacqmin_jfm_2000} and can enforce the thermodynamical consistency of the models. Although many computational studies have addressed the unfavourable fluid displacement problem in a rigid capillary tube \cite{suo_wrr_2024, gao_etal_jfm_2019, esmaeilzadeh_etal_pre_2020}, there are currently no reported computational methods for studying this fluid displacement phenomenon in a soft capillary tube. This represents a critical gap in our understanding, as a soft tube introduces fundamentally different physics due to the coupling between fluids motion and the tube deformability. In a soft tube, the interplay of capillary forces, viscous forces, and elastic forces leads to complex interfacial dynamics that can significantly alter the onset of interfacial instabilities and the bubble pinch-off behavior previously known in a rigid tube. A computational framework for studying this problem is important to better understand how the tube’s deformability affects the interfacial dynamics, the onset of interfacial instability, and bubble pinch-off. Findings from such numerical simulations can guide future experiments in a soft capillary tube. This problem will also prove important in many practical applications such as soft porous media, bio-microfluidics, and pressure-actuated soft robotics, where interactions between fluids and deformable solids are paramount. More broadly, our study lays the groundwork for achieving a better control and manipulation of immiscible fluids in soft, confined geometries.

Here, we present a novel computational fluid–structure interaction (FSI) method to simulate the unfavourable fluid-fluid displacement problem in a soft capillary tube. We employ a phase-field model for the fluids and a nonlinear solid model capable of capturing large solid deformations. 
Due to the strong coupling between fluid flow and solid deformation, we adopt a monolithic framework rather than a partitioned approach. Our FSI formulation uses a boundary-fitted approach, which enables accurate computation of forces at the fluid–solid interface. For spatial discretization, we use Isogeometric Analysis, which allows basis functions with controllable inter-element continuity---enhancing both accuracy and stability of our numerical simulations. Our framework incorporates a combination of complex multiphysics phenomena that has not been studied before: two-component fluid flow with topological changes, nonlinear solid deformation, surface tension effects, dynamic wetting, and contact line motion. By systematically varying the tube’s elasticity and the inlet capillary number, we develop a comprehensive understanding of when and how interfacial instability and bubble pinch-off occur in a soft capillary tube, for both imbibition and drainage cases. 

Our simulations reveal that the tube deformability suppresses or delays the onset of interfacial instability and bubble pinch-off, compared to the rigid tube scenario for all the values of wettability that we studied. However, the mechanisms whereby interfacial instability is delayed in wetting and non-wetting systems are different. While in a wetting system, the stabilizing effect of the tube's compliance can be simply explained by the expansion of the tube's cross section that reduces the fluid velocity, in a non-wetting case, the role of the tube's compliance is more complex because there is a stabilizing mechanism and a de-stabilizing mechanism. Our detailed quantitative analysis, will show that even in the non-wetting case, the stabilizing effect dominates and always delays or suppresses the instability.  

\section{Governing equations}

We consider a binary fluid composed of two immiscible constituents: a liquid with specific wettability properties and air. We assume that the two immiscible constituents flow in a soft capillary tube. In what follows, we use the superscript $\hat{}$ to denote dimensional variables and we omit that superscript to indicate the corresponding dimensionless variables. 

\subsection{Kinematics}

Let $\spatdomu$ and $\refdomu$ be sets in $\rspace^d$, where $d$ is the number of spatial dimensions and $\hat{t}$ represents time. To avoid distracting mathematical details, we will treat sets as open or closed as convenient. The sets $\spatdomu$ and $\refdomu$ refer to the current and reference configurations occupied by a deformable continuum body, respectively. We assume that $\refdomu$ is fixed in time and parameterized by the reference coordinates $\hat{\bm{X}}$. We describe the motion of the deformable body by a function $\hat{\map}$ at time $\hat{t}$, defined as $\hat{\map}(\cdot, \hat{t}) : \refdomu \longmapsto \spatdomu = \hat{\map}(\refdomu, \hat{t})$, such that $\hat{\bm{X}} \longmapsto \hat{\bm{x}} = \hat{\map}(\hat{\bm{X}}, \hat{t}) \ \forall \hat{\bm{X}} \in \refdomu$, where $\hat{\bm{x}}$ denotes the coordinates of the spatial domain. We express the referential displacement by $\hat{\disps}(\hat{\bm{X}}, \hat{t}) = \hat{\map}(\hat{\bm{X}}, \hat{t}) - \hat{\bm{X}}$, and the referential velocity by $\hat{\velf} = \frac{\partial \hat{\map}}{\partial \hat{t}}$. We define the deformation gradient by $\defgrad = \frac{\partial \hat{\map}}{\partial \hat{\bm{X}}}$, and the Jacobian determinant by $J = \det \defgrad$. In what follows, we use subscripts in the definition of spatial and time derivatives. The subscript $\hat{\bm{X}}$ in $\left. \frac{\partial \hat{\disps}}{\partial \hat{t}} \right|_{\hat{\bm{X}}}$ indicates that we compute the time derivative by keeping $\hat{\bm{X}}$ fixed. When we do not specify any subscript in the time derivative, the time derivative is computed by holding $\hat{\bm{x}}$ fixed. Similarly, for spatial derivatives, the subscript $\hat{\bm{X}}$ in $\nabla_{\hat{\bm{X}}} \hat{\disps}$, for example, indicates that we compute the spatial derivative with respect to $\hat{\bm{X}}$. When we do not specify any subscript in the spatial derivative, the derivative is computed with respect to $\hat{\bm{x}}$.

In the fluid-structure interaction (FSI) problem we consider here, the spatial domain $\spatdomu$ is decomposed into two sets, $\spatdomfu$ and $\spatdomsu$, such that $\spatdomu = \spatdomfu \cup \spatdomsu$ and $\spatdomfu \cap \spatdomsu = \emptyset$. The sets $\spatdomfu$ and $\spatdomsu$ represent the current configurations of the fluid and solid, respectively. We apply a similar decomposition to the referential domain $\refdomu = \refdomfu \cup \refdomsu$, where $\refdomfu$ and $\refdomsu$ denote the reference configurations of the fluid and solid, respectively. We denote the fluid-solid interface in the spatial domain as $\sfintu$, fluid-solid interface in the referential domain as $\sfintrefu$, external fluid boundary in the spatial domain excluding the fluid-solid interface as $\fintu$ and the external solid boundary in the referential domain excluding the fluid-solid interface as $\sintrefu$.

\subsection{Governing equations of the coupled fluid-solid interaction problem}
\label{sec:strongform}

\subsubsection{Fluid mechanics}

We describe the dynamics of liquid and air using the Navier-Stokes-Cahn-Hilliard (NSCH) equations---a phase-field model for describing the flow of two immiscible fluids \cite{abels_etal_2012, gomez_etal_frontiers_2023, khanwale_cpc_2022}. We assume both the liquid and air are individually incompressible, and share the same velocity field. We define the free energy per unit volume of liquid and air as 
\begin{equation*}
    \glenergy = \frac{\hat{\lambda}}{4 \eps^2} \lbr 1 - \phase^2 \rbr^2 + \frac{\hat{\lambda}}{2}|\hat{\nabla} \phase|^2,
\end{equation*}
where $\hat{\lambda} = \frac{3}{2 \sqrt{2}} \fluidsf \eps$ \cite{dong_jcp_2014, dong_shen_jcp_2012} is the mixing energy per unit volume of liquid and air, $\eps$ is the diffuse interface length scale, $\fluidsf$ is the surface tension at the liquid-air interface and $\phase \in [-1,1]$ is a phase-field denoting the volume fraction difference between the liquid and air. It follows from the definition of $c$ that $\phase = 1$ represents the liquid and $\phase = -1$ represents air. The governing equations of the fluids include the mass balance and the linear momentum balance equations written in the current configuration. We neglect the relative mass diffusive flux term from the momentum equation in the originally derived thermodynamically consistent NSCH equations \cite{abels_etal_2012, khanwale_cpc_2022}. We have verified from our simulations that neglecting this term has a negligible impact on our results. \sbr{We also neglect gravitational forces in our governing equations because the reference length scale of our simulation is smaller than the capillary length $\sqrt{\frac{\fluidsf}{\densfl \hat{g}}}$. Here, $\densfl$ is the true density of the liquid (i.e., mass of the liquid over the volume occupied by the liquid) and $\hat{g}$ is the acceleration due to gravity}.

\subsubsection{Solid mechanics}

We assume the solid to be hyperelastic, isotropic and homogeneous. We describe the dynamics of the solid using the linear momentum balance equation written in the reference configuration. To describe the material behavior of the solid, we use a neo-Hookean constitutive model \cite{simo_2006_book}. The strain energy per unit volume of undeformed configuration of the solid is given by 
\begin{equation*}
    \hat{W} = \frac{\hat{\kappa}}{2} \left(\frac{1}{2}\left(J^2 - 1\right) - \ln{J}\right) + \frac{\hat{G}}{2}\left(J^{-\sfrac{2}{3}}\tr\left({\bm{C}}\right) - 3 \right),
\end{equation*}
where $\hat{\kappa}$ and $\hat{G}$ are the bulk and shear moduli of the solid and $\bm{C} = \defgrad^T\defgrad$ is the right Cauchy-Green deformation tensor. The Young's modulus and the Poisson's ratio of the solid are defined from $\hat{G}$ and $\hat{\kappa}$ as $\hat{E} = \frac{9 \hat{\kappa} \hat{G}}{3 \hat{\kappa} + \hat{G}}$ and $\nu = \frac{3\hat{\kappa} - 2 \hat{G}}{2 \left(3 \hat{\kappa} + \hat{G} \right)}$, respectively.

\subsubsection{Fluid-structure interaction problem}

We simplify our three-dimensional FSI problem by assuming axisymmetry. This assumption reduces the computational cost of our simulations, while preserving the essential physics of the original three-dimensional problem. In what follows, we denote any vector written in spatial coordinates as $\bm{a} = \lbr a_{{r}}, a_{{z}} \rbr$ and any tensor written in spatial coordinates as $\bm{A} = \lbr A_{ij} \rbr$, where $i, j = {r}, \vartheta, {z}$. Here, ${r}$ and ${z}$ denote the radial and axial directions of the spatial domain, respectively, while $\vartheta$ denotes the azimuthal direction. Similarly, in reference coordinates, we denote any vector as $\bm{b} = \lbr b_{{R}}, b_{{Z}} \rbr$, and any tensor as $\bm{B} = \lbr B_{IJ} \rbr$, where $I, J = {R}, \vartheta, {Z}$. Here, ${R}$ and ${Z}$ denote the radial and axial directions of the referential domain, respectively. At the level of differential operators, the axisymmetric assumption is realized using $\hat{\nabla} = \lbr \frac{\partial}{\partial \xrd},  \frac{\partial}{\partial \xzd} \rbr$, $\hat{\nabla} \cdot \bm{a} = \frac{1}{\xrd}\frac{\partial}{\partial \xrd} \lbr \xrd a_{{r}} \rbr + \frac{\partial a_{{z}}}{\partial \xzd}$, $\hat{\nabla} \cdot \bm{A} = \lbr \frac{\partial A_{{r}{r}}}{\partial \xrd} + \frac{1}{\xrd} \lbr A_{{r}{r}} - A_{\vartheta \vartheta} \rbr + \frac{\partial A_{{z}{r}}}{\partial \xzd}, \frac{\partial A_{{r}{z}}}{\partial \xrd} + \frac{\partial A_{{z}{z}}}{\partial \xzd} + \frac{A_{{r}{z}}}{\xrd} \rbr$ and $\hat{\Delta} = \frac{1}{\xrd} \frac{\partial}{\partial \xrd} + \frac{\partial}{\partial \xrd}\frac{\partial}{\partial \xrd} + \frac{\partial}{\partial \xzd}\frac{\partial}{\partial \xzd}$, where $\xrd$ and $\xzd$ refer to the radial and axial spatial coordinates. We define other differential vector and tensor operators accordingly. In addition, we define the differential operators that involve $\hat{\nabla}_{\hat{\bm{X}}}$ as a function of the radial and axial reference coordinates $\Xrd$ and $\Xzd$, respectively. 


We non-dimensionalize the variables with reference length $\hat{L}_r$, reference time $\hat{t}_r = \frac{\viscfl \hat{L}_r}{\fluidsf}$ and reference mass $\hat{m}_r = \densfl \hat{L}_r^3$, respectively. Here, $\viscfl$ is the dynamic viscosity of the liquid. We use the following dimensionless numbers in our governing equations: a) Ohnesorge number of the liquid $\ohno = \frac{\viscfl}{\sqrt{\densfl \fluidsf \hat{L}_r}}$ which is defined as the ratio of inertio-capillary to the inertio-viscous time scale; b) viscosity ratio of liquid to air $\tilde{\viscf} = \frac{\viscfl}{\viscfa}$, where $\viscfa$ is the dynamic viscosity of air; c) density ratio of liquid to air $\tilde{\densf} = \frac{\densfl}{\densfa}$, where $\densfa$ is the true density of air; d) Cahn number $\cahnno = \frac{\eps}{\hat{L}_r}$; e) P\'eclet number $\pecno = \frac{\hat{L}_r^2}{\hat{M} \viscfl}$ which is the ratio of advection to diffusion, where $\hat{M}$ is the positive mobility coefficient; \sbr{f) elastocapillary number which quantifies the strength of elastocapillary effects given by $\zeta = \frac{\hat{E} \hat{h}^3}{\fluidsf \hat{L}_r^2}$, where $\hat{h}$ is the radial solid thickness, g) the Poisson's ratio $\nu$ and h) capilloelastic number $\chi = \frac{\fluidsf}{\hat{E} \hat{L}_r}$.} 

We define subsets of $\fint$ such that $\fint = \fintdir \cup \fintneu \cup \fintaxi$, where Dirichlet fluid velocity boundary conditions are imposed on $\fintdir$, traction-free boundary conditions are imposed on $\fintneu$, and $\fintaxi$ coincides with the axis of symmetry. Similarly, we define subsets of $\sintref$ such that $\sintref = \sintdir \cup \sintneu$, where we impose zero solid displacements on $\sintdir$ and traction-free boundary conditions on $\sintneu$. Given initial conditions for the fluid velocity ($\velf_0$), solid displacement ($\disps_0$) and the phase-field ($\phase_0$), we define the dimensionless governing equations for our FSI problem as follows: find $p : \spatdomf \times \left(0, T \right] \longmapsto \rspace$, $\velf : \spatdomf \times \left(0, T \right] \longmapsto \rspace^d$, $\phase : \spatdomf \times \left(0, T \right] \longmapsto \left[-1, 1\right]$, $\mu : \spatdomf \times \left(0, T \right] \longmapsto \rspace$ and $\disps : \refdoms \times \left( 0, T \right] \longmapsto \rspace^d$ such that
\begin{subequations}
    \begin{alignat}{2}
        & \frac{1}{\xr} \frac{\partial}{\partial \xr} (\xr \vr) + \frac{\partial \vz}{\partial \xz} = 0 && \quad \text{in} \ \spatdomf \times (0, T] \label{eqn:continuity_eqn} \\
        &\begin{aligned}
            \rho &\left(\dt \vr + \vr \frac{\partial \vr}{\partial \xr} + \vz \frac{\partial \vr}{\partial \xz} \right) 
            - \frac{\partial \sigmafrr}{\partial \xr}
            - \frac{1}{\xr} \lbr \sigmafrr - \sigmaftt \rbr - \frac{\partial \sigmafrz}{\partial \xz} = 0\\ 
        \end{aligned} && \quad \text{in} \ \spatdomf \times (0, T] \label{eqn:momentum_eqn_r} \\
        &\begin{aligned}
            \rho &\left(\dt \vz + \vr \frac{\partial \vz}{\partial \xr} + \vz \frac{\partial \vz}{\partial \xz} \right) 
            - \frac{\partial \sigmafrz}{\partial \xr} - \frac{1}{\xr}\sigmafrz - \frac{\partial \sigmafzz}{\partial \xz} = 0 \\ 
        \end{aligned} && \quad \text{in} \ \spatdomf \times (0, T] \label{eqn:momentum_eqn_z} \\
        & \dt \phase + \vr \frac{\partial \phase}{\partial \xr} + \vz \frac{\partial \phase}{\partial \xz} = \frac{1}{\pecno} \Bigg( \frac{1}{\xr} \frac{\partial}{\partial \xr} \lbr \xr \frac{\partial \mu}{\partial \xr} \rbr  + \frac{\partial^2 \mu}{\partial \xz^2} \Bigg)
        \quad && \quad \text{in} \ \spatdomf \times (0, T] \label{eqn:phfield_eqn} 
        \\
        & \mu = \frac{3}{2 \sqrt{2}} \Bigg[ \frac{1}{\cahnno} \phase \lbr \phase^2 - 1 \rbr - \cahnno \Bigg( \frac{1}{\xr} \frac{\partial}{\partial \xr} \lbr \xr \frac{\partial \phase}{\partial \xr} \rbr + \frac{\partial^2 \phase}{\partial \xz^2} \Bigg) \Bigg] && \quad \text{in} \ \spatdomf \times \left( 0, T \right] 
        \label{eqn:chempot_eqn}        
        \\        
        & \left. \dt\dt \ur \right|_{\bm{X}} = \frac{\partial \pkonerr}{\partial \Xr} + \frac{1}{\Xr} \lbr \pkonerr - \pkonett \rbr + \frac{\partial \pkonerz}{\partial \Xz} && \quad \text{in} \ \refdoms \times \left( 0, T \right] 
        \label{eqn:momentum_eqn_solid_1}
        \\
        & \left. \dt\dt \uz \right|_{\bm{X}} = \frac{\partial \pkonerz}{\partial \Xr} + \frac{1}{\Xr}\pkonerz + \frac{\partial \pkonezz}{\partial \Xz} && \quad \text{in} \ \refdoms \times \left( 0, T \right] 
        \label{eqn:momentum_eqn_solid_2}     
    \end{alignat}
    \label{eq:strongform}
\end{subequations}
$\!\!\!$where $p$ is the fluid pressure, $\mu$ is the chemical potential, $\left[0, T \right]$ is the time interval of interest, $\densf$ is the  density of the fluid mixture, $\dt$ denotes partial time differentiation, $\bm{\sigma}^f$ is the fluid Cauchy stress tensor, $\disps$ is the solid displacement, and $\pkone$ is the first Piola-Kirchhoff stress tensor. In Eqs.~\eqref{eqn:momentum_eqn_r} and \eqref{eqn:momentum_eqn_z}, $\densf = \frac{1}{2} \lbr 1 + \frac{1}{\tilde{\densf}}\rbr + \frac{1}{2} \lbr 1 - \frac{1}{\tilde{\densf}}\rbr \phase$. The Cauchy stress of the fluid system is 
\begin{equation*}
    \bm{\sigmaf} = - p \bm{I} + 2 \ohno^2 \viscf(\phase)  \ \nabla^{s} \bm{v} - \frac{3}{2\sqrt{2}} \ \cahnno \ \ohno^2 \ \nabla c \otimes \nabla c.
\end{equation*}
Here, $\viscf (\phase) = \frac{1}{2}\lbr 1 + \frac{1}{\tilde{\viscf}} \rbr + \frac{1}{2}\lbr 1 - \frac{1}{\tilde{\viscf}} \rbr c$ and $\nabla^{s}$ is the symmetrization of the spatial gradient operator $\nabla$. In Eqs.~\eqref{eqn:momentum_eqn_solid_1} and \eqref{eqn:momentum_eqn_solid_2}, the first Piola-Kirchhoff stress tensor is obtained as $\pkone = 2 \defgrad \frac{\partial W}{\partial \bm{C}}$, leading to
\begin{equation*}
    \bm{\pkone} =  
\frac{\ohno^2}{3 \chi \lbr 1 - 2\nu \rbr} \frac{\densfl}{\denss} \bm{\defgrad} J^{-\sfrac{2}{3}}\Bigl(\bm{I} - \frac{1}{3} \tr\lbr {\bm{C}} \rbr \bm{C}^{-1}\Bigr) + \frac{\ohno^2}{6 \chi \lbr 1 + \nu \rbr} \frac{\densfl}{\denss} \bm{\defgrad} \Big( J^{2} - 1 \Big) \bm{C}^{-1}.
\end{equation*}
Here, $\denss$ is the mass density of the solid in the reference configuration. Comparing our fluid governing equations with those in \cite{khanwale_cpc_2022}, we scale $p$ differently here, specifically with $\frac{\hat{m}_r}{\hat{L}_r \hat{t}_r^2}$. 

The boundary and interface conditions for our FSI problem are 
\begin{subequations}
    \begin{alignat}{2}
        & \velf = \velf_\mathrm{in}  \quad && \quad \text{on} \ \fintdir \ \times \left[0, T \right]
        \label{eqn:dirichlet_vel_extbdary}
        \\
        & \sigmaf \normf = 0 \quad && \quad \text{on} \ \fintneu \ \times \left[0, T \right] \label{eqn:tractionfree_velbdary} \\
        & \bm{e}_r \cdot \sigmaf \normf = 0 \quad &&
        \begin{aligned} & \quad \text{on} \ \fintaxi \ \times \left[0, T \right] \label{eqn:tractionfree_axidbdary} 
        \end{aligned} \\      
        & \velf \cdot \normf = 0 \quad && \quad \text{on} \ \fintaxi \ \times \left[0, T \right] \label{eqn:normalvel_axidbdary} \\       
        & \normf \cdot \nabla \phase = 0 \quad && \quad \text{on} \ \fint \ \times \left[ 0, T \right] 
        \label{eqn:freeflux_phfield_fluidbdary}
        \\        
        & \normf \cdot \nabla \mu = 0 \quad && \quad \text{on} \ \fint \ \times \left[ 0, T \right]
        \label{eqn:chempot_fluidbdary}
        \\        
        & \disps = 0  \quad && \quad \text{on} \ \sintdir \ \times \left[0, T \right] 
        \label{eqn:soliddisp_solidbdary}
        \\        
        & \pkone \normsref = 0  \quad && \quad \text{on} \ \sintneu \ \times \left[0, T \right] 
        \label{eqn:tractionfree_solidbdary}
        \\
        & \velf - \vels \circ \map^{-1} = 0 \quad && \quad \text{on} \ \sfint \ \times \left[0, T \right] 
        \label{eqn:fluidvel_soliddisp_fsibdary}
        \\
        & \sigmaf \normsf - \sigmas \normsf = 0 \quad && \quad \text{on} \ \sfint \ \times \left[ 0, T \right] \label{eqn:tractionbalance_fsibdary} \\
        & \normsf \cdot \nabla \phase = \frac{2 \sqrt{2}}{3\cahnno} h_w \lbr \phase, \vr, \vz \rbr \quad && \quad \text{on} \ \sfint \ \times \left[0, T \right] 
        \label{eqn:wettingbalance_fsibdary}
        \\
        & \normsf \cdot \nabla \mu = 0 \quad && \quad \text{on} \ \sfint \ \times \left[ 0, T \right]
        \label{eqn:chempot_fsibdary}  
    \end{alignat}
    \label{eq:strongform_bcs}
\end{subequations}
$\!\!\!$where $\velf_\mathrm{in}$ denotes the Dirichlet fluid velocity boundary condition on $\fintdir$. On $\fintaxi$, we impose a zero tangential traction condition (see Eq.~\eqref{eqn:tractionfree_axidbdary}), zero normal fluid velocity (see Eq.~\eqref{eqn:normalvel_axidbdary}), a neutral wettability condition (see Eq.~\eqref{eqn:freeflux_phfield_fluidbdary}) and a zero diffusive flux (see Eq.~\eqref{eqn:chempot_fluidbdary}). On the fluid-solid interface $\sfint$, we impose kinematic compatibility of the fluid and solid velocities (see Eq.~\eqref{eqn:fluidvel_soliddisp_fsibdary}), balance of the fluid and solid tractions (see Eq.~\eqref{eqn:tractionbalance_fsibdary}), a dynamic wettability condition (see Eq.~\eqref{eqn:wettingbalance_fsibdary}) and zero diffusive flux (see Eq.~\eqref{eqn:chempot_fsibdary}). In Eq.~\eqref{eqn:wettingbalance_fsibdary}, $h_w$ is a function of $\phase, \vr$ and $\vz$ that determines the affinity of the fluid in contact with the solid. The rest of the notation in Eq.~\eqref{eq:strongform_bcs} is as follows: $\normf$ is the unit outward normal vector to $\fint$, $\normsref$ is the unit outward normal vector to $\sintref$, $\normsf$ is the unit normal vector to $\sfint$ in the direction from fluid to solid, and $\sigmas$ is the solid Cauchy stress tensor. In Eq.~\eqref{eqn:tractionbalance_fsibdary}, $\sigmas = J^{-1} \pkone \defgrad^{T}$. In Eq.~\eqref{eqn:wettingbalance_fsibdary}, $h_w$ is given by
\begin{equation}
    h_w \lbr \phase, \vr, \vz \rbr = - \dw \big( \dt \phase + \velf \cdot \nabla \phase \big) + \frac{3}{4} \lbr 1 - \phase^2 \rbr \cos \theta,
    \label{eqn:dynamic_contactangle_law}
\end{equation}
\noindent where $\dw$ is a non-negative dynamic wall mobility coefficient that accounts for wall relaxation and bulk diffusion near the fluid-solid interface and $\theta$ is the equilibrium contact angle made by the air-liquid interface with the solid. Eqs.~\eqref{eqn:wettingbalance_fsibdary} and ~\eqref{eqn:dynamic_contactangle_law} are the dimensionless form of the dynamic wettability condition used in \cite{yue_feng_pof_2011}. \sbr{In the present work, we assume that $\dw$ is only a function of $\theta$. As we show later in Sec.~\ref{sec:results}, this dependence of $\dw$ on $\theta$ yields good agreement between our simulations and previous experiments.} Eq.~\eqref{eqn:dynamic_contactangle_law} shows that an increase in the value of $\dw$ increases the deviation from the equilibrium contact angle.

\section{Numerical formulation}


We solve the governing equations using a body-fitted fluid-structure algorithm, similar to the approach used in \cite{bueno_2018b, bhopalam_cmame_2022}. Following \cite{donea_2004}, we rewrite Eqs.~\eqref{eqn:continuity_eqn}--\eqref{eqn:phfield_eqn} in the Arbitrary Lagrangian-Eulerian (ALE) form. The ALE equations involve $\vrm$ and $\vzm$, which are the radial and axial components of the fluid mesh velocity $\widetilde{\velf}$, respectively. For our body-fitted fluid-structure algorithm, we update the fluid mesh by solving successive fictitious linear elasticity problems \cite{wick_2011}.

\subsection{Continuous problem in variational form}

We derive the weak formulation of Eq.~\eqref{eq:strongform} with the fluid governing equations written in the ALE description. Let us introduce the trial solution spaces $\bm{\mathcal{X}}^f = \bm{\mathcal{X}}^f(\spatdomf)$, $\bm{\mathcal{X}}^s = \bm{\mathcal{X}}^s(\refdoms)$ and $\bm{\mathcal{X}}^m$ be defined on the fluid mesh domain. Let $\bm{U}^f = \{p, \vr, \vz, \phase, \mu\}$ be a typical member of $\bm{\mathcal{X}}^f$, $\bm{U}^s = \{\ur, \uz\}$ be a typical member of $\bm{\mathcal{X}}^s$ and $\bm{U}^m$ be a typical member of $\bm{\mathcal{X}}^m$. The weight function spaces $\bm{\mathcal{Y}}^f = \bm{\mathcal{Y}}^f(\spatdomf)$, $\bm{\mathcal{Y}}^s = \bm{\mathcal{Y}}^s(\refdoms)$ and $\bm{\mathcal{Y}}^m$ defined on the fluid mesh domain are identical to $\bm{\mathcal{X}}^f$, $\bm{\mathcal{X}}^s$ and $\bm{\mathcal{X}}^m$ respectively, but all restrictions on the Dirichlet boundary are homogeneous. In what follows, we denote standard $\mathcal{L}^2$ inner products over $\spatdomf$, $\refdoms$, and $\sfint$ as $\lbr \cdot, \cdot \rbr_{\spatdomf}$, $\lbr \cdot, \cdot \rbr_{\refdoms}$, and $\lbr \cdot, \cdot \rbr_{\sfint}$, respectively. Before integrating over $\spatdomf$ and $\refdoms$, we multiplied the governing equations by $\xr$ and $\Xr$ respectively. We define the weak formulation of the coupled fluid-structure interaction problem as follows: find $\bm{U}^f = \{p, \vr, \vz, \phase, \mu\} \in \bm{\mathcal{X}}^f$, $\bm{U}^s = \{\ur, \uz \} \in \bm{\mathcal{X}}^s$ and $\bm{U}^m \in \bm{\mathcal{X}}^m$ such that $\forall \bm{W}^f = \{\wp, \wvr, \wvz, \wc, \wmu\} \in \bm{\mathcal{Y}}^f$, $\forall \bm{W}^s = \{\wur, \wuz \} \in \bm{\mathcal{Y}}^s$ and $\forall \bm{W}^m \in \bm{\mathcal{Y}}^m$,
\begin{equation}
    B^f(\bm{W}^f, \bm{U}^f) + B^s(\bm{W}^s, \bm{U}^s) + B^m(\bm{W}^m, \bm{U}^m) = \lbr \xr \wmu, h_w \rbr_{\sfint},
    \label{eqn:weakform_cont}
\end{equation}
\noindent where
\begin{equation}
\begin{aligned}
B^f(& \bm{W}^f, \bm{U}^f) = \lbr \xr \wp, \frac{1}{\xr} \frac{\partial}{\partial \xr} (\xr \vr) + \frac{\partial \vz}{\partial \xz} \rbr_{\spatdomf} \\
& + \lbr \xr \wvr, \densf \lbr \phase \rbr \lbr \left. \dt \vr \right|_{\xm} + \lbr \vr - \vrm \rbr \frac{\partial \vr}{\partial \xr} + \lbr \vz - \vzm \rbr \frac{\partial \vr}{\partial \xz} \rbr \rbr - \lbr \xr \frac{\partial \wvr} {\partial \xr}, p \rbr_{\spatdomf} - \lbr \xr \wvr, p \rbr_{\spatdomf} \\
& + \lbr \xr \frac{\partial \wvr}{\partial \xr}, 2 \ohno^2 \viscf(\phase) \frac{\partial \vrh}{\partial \xr} \rbr_{\spatdomf} + \lbr \xr \frac{\partial \wvr}{\partial \xz}, \ohno^2 \viscf(\phase) \lbr \frac{\partial \vr}{\partial \xz} + \frac{\partial \vz}{\partial \xr} \rbr \rbr_{\spatdomf} + \lbr \wvr, 2 \ohno^2 \viscf(\phase) \frac{\vr}{\xr} \rbr_{\spatdomf} \\
& - \lbr \xr \frac{\partial \wvr}{\partial \xr}, \frac{3}{2\sqrt{2}} \ \cahnno \ \ohno^2 \frac{\partial \phase}{\partial \xr} \frac{\partial \phase}{\partial \xr} \rbr_{\spatdomf} - \lbr \xr \frac{\partial \wvr}{\partial \xz}, \frac{3}{2\sqrt{2}} \ \cahnno \ \ohno^2 \frac{\partial \phase}{\partial \xr} \frac{\partial \phase}{\partial \xz} \rbr_{\spatdomf} \\
& + \Bigg( \xr \wvz,  \rho \lbr \phase \rbr \lbr \left. \dt \vz \right|_{\xm} + \lbr \vr - \vrm \rbr \frac{\partial \vz}{\partial \xr} + \lbr \vz - \vzm \rbr \frac{\partial \vz}{\partial \xz} \rbr \Bigg)_{\spatdomf} - \lbr \xr \frac{\partial \wvz} {\partial \xz}, p \rbr_{\spatdomf} \\
& + \lbr \xr \frac{\partial \wvz}{\partial \xz}, 2 \ohno^2 \viscf(\phase) \frac{\partial \vz}{\partial \xz} \rbr_{\spatdomf} + \lbr \xr \frac{\partial \wvz}{\partial \xr}, \ohno^2 \viscf(\phase) \lbr \frac{\partial \vr}{\partial \xz} + \frac{\partial \vz}{\partial \xr} \rbr \rbr_{\spatdomf} \\
& - \lbr \xr \frac{\partial \wvz}{\partial \xr}, \frac{3}{2\sqrt{2}} \ \cahnno \ \ohno^2 \frac{\partial \phase}{\partial \xz} \frac{\partial \phase}{\partial \xr} \rbr_{\spatdomf} - \lbr \xr \frac{\partial \wvz}{\partial \xz}, \frac{3}{2\sqrt{2}} \ \cahnno \ \ohno^2 \frac{\partial \phase}{\partial \xz} \frac{\partial \phase}{\partial \xz} \rbr_{\spatdomf} \\
& + \lbr \xr \wc, \left. \dt \phase \right|_{\xm} + \lbr \vr - \vrm \rbr \frac{\partial \phase}{\partial \xr} + \lbr \vz - \vzm \rbr \frac{\partial \phase}{\partial \xz} \rbr_{\spatdomf} + \lbr \xr \frac{\partial \wc}{\partial \xr}, \frac{1}{\pecno} \frac{\partial \mu}{\partial \xr} \rbr_{\spatdomf} + \lbr \xr \frac{\partial \wc}{\partial \xz}, \frac{1}{\pecno} \frac{\partial \mu}{\partial \xz} \rbr_{\spatdomf} \\
& + \lbr \xr \wmu, \mu \lbr \phase \rbr - \frac{3}{2 \sqrt{2}\cahnno} \phase \lbr \phase^2 - 1 \rbr \rbr_{\spatdomf} - \lbr \xr \frac{\partial \wmu}{\partial \xr}, \frac{3}{2 \sqrt{2}}\cahnno \frac{\partial \phase}{\partial \xr} \rbr_{\spatdomf} - \lbr \xr \frac{\partial \wmu}{\partial \xz}, \frac{3}{2 \sqrt{2}} \cahnno \frac{\partial \phase}{\partial \xz} \rbr_{\spatdomf}, \\
\end{aligned}
\label{weakform_cont_fluid}
\end{equation}
\begin{equation}
\begin{aligned}
    B^s(& \bm{W}^s, \bm{U}^s) = \Big(\Xr \wur, \left. \dt\dt \ur \right|_{\bm{X}} \Big)_{\refdoms} + \lbr \Xr \frac{\partial \wur}{\partial \Xr}, \pkonerr \rbr_{\refdoms} + \lbr \Xr \frac{\partial \wur}{\partial \Xz}, \pkonerz \rbr_{\refdoms} \\ 
    & + \Big( \wur, \pkonett \Big)_{\refdoms} + \Big(\Xr \wuz, \left. \dt\dt \uz \right|_{\bm{X}} \Big)_{\refdoms} + \lbr \Xr \frac{\partial \wuz}{\partial \Xr}, \pkonerz \rbr_{\refdoms} + \lbr \Xr \frac{\partial \wuz}{\partial \Xz}, \pkonezz \rbr_{\refdoms},
\end{aligned}
\label{weakform_cont_solid}
\end{equation}
\noindent and $B^m \lbr \cdot, \cdot \rbr$ is the weak formulation of the mesh motion subproblem; see \cite{bazilevs_fsirev_2008}. Eqs.~\eqref{eqn:weakform_cont}-\eqref{weakform_cont_solid}, weakly enforce the boundary conditions given by Eqs.~\eqref{eqn:tractionfree_velbdary}, ~\eqref{eqn:tractionfree_axidbdary}, ~\eqref{eqn:freeflux_phfield_fluidbdary}, ~\eqref{eqn:chempot_fluidbdary}, ~\eqref{eqn:tractionfree_solidbdary} and ~\eqref{eqn:chempot_fsibdary}. To enforce the traction balance as in Eq.~\eqref{eqn:tractionbalance_fsibdary}, we take $\wvr|_{\sfint} = \wur \circ \map^{-1}|_{\sfint}$ and $\wvz|_{\sfint} = \wuz \circ \map^{-1}|_{\sfint}$. The term on the right hand side of Eq.~\eqref{eqn:weakform_cont} follows from the wetting condition in Eq.~\eqref{eqn:wettingbalance_fsibdary}. 

\subsection{Semi-discrete variational formulation of the coupled fluid-structure interaction problem} 

For the spatial discretization of Eq.~\eqref{eqn:weakform_cont}, we use the Galerkin method for the solid and mesh motion sub-problems, and the Variational Multiscale method (VMS) \cite{hughes_cmame_1998, hughes_etal_2018} for the fluid sub-problem. We define finite-dimensional approximations of the trial and weight function spaces, denoted by $\bm{\mathcal{X}}^{s,h}$, $\bm{\mathcal{X}}^{m,h}$ and $\bm{\mathcal{Y}}^{s,h},\, \bm{\mathcal{Y}}^{m,h}$ such that $\bm{\mathcal{X}}^{s,h} \subset \bm{\mathcal{X}}^{s}$, $\bm{\mathcal{Y}}^{s,h} \subset \bm{\mathcal{Y}}^{s}$, $\bm{\mathcal{X}}^{m,h} \subset \bm{\mathcal{X}}^{m}$ and $\bm{\mathcal{Y}}^{m,h} \subset \bm{\mathcal{Y}}^{m}$. In the VMS approach, we decompose $\xf$ and $\yf$ into coarse scale and subgrid scale subspaces via a multiscale direct-sum decomposition, i.e., $\xf = \xfh \oplus \xfd$ and $\yf = \yfh \oplus \yfd$, where the finite-dimensional functional space associated with the coarse scale is denoted by the superscript \quotes{f, h} while the infinite-dimensional functional space associated with the subgrid scale is denoted by the superscript \quotes{f, '}. The split and notation also apply to the unknowns $\bm{U}^f$, the weight functions $\bm{W}^f$ and their individual components. Following \cite{khanwale_cpc_2022, bazilevs_2013}, we choose $\phase' = \mu' = \wpd = \wvrd = \wvzd = \wcd = \wmud = 0$. We model the subgrid scale trial functions of velocity and pressure using residual-based approximations as follows, 
\begin{equation}
    \begin{aligned}
        &  \densf \vr' = -\taum R_{mr} \lbr \ph, \vrh, \vzh, \phaseh, \muh; \vrmh, \vzmh \rbr, \\ 
         & \densf \vz' = -\taum R_{mz} \lbr \ph, \vrh, \vzh, \phaseh, \muh; \vrmh, \vzmh \rbr, \\
         & p' = -\densf \tauc R_{co} \lbr \vrh, \vzh \rbr
    \end{aligned}
     \label{eqn:subgridscale_models}
\end{equation}
\noindent  where $R_{mr}$ and $R_{mz}$ are the residuals of the momentum equations in ALE description while $R_{co}$ is the residual of the continuity equation in ALE description, respectively. The stabilization parameters in Eq.~\eqref{eqn:subgridscale_models} are defined as
\begin{equation}
    \taum = \Bigg(\frac{4}{(\Delta t)^2} + ({\bm{v}}^h - \tilde{\bm{v}}^h) \cdot \bm{G} ({\bm{v}}^h - \tilde{\bm{v}}^h) + C_I \left(\frac{\viscf \lbr \phase^h \rbr}{\densf \lbr \phase^h \rbr}\right)^2 \bm{G}:\bm{G} \Bigg)^{-1/2}, \quad 
    \tauc = \frac{1}{\tr(\bm{G})\taum},
\end{equation}
\noindent where $\Delta t$ is the discrete time step size, $C_I$ is a positive constant derived from an appropriate element-wise inverse estimate \cite{johnson_book_2009} and $\bm{G}$ is the element metric tensor \cite{bazilevs_2013}, i.e., $G_{ij} = \sum_{k = 1}^d \frac{\partial \zeta_k}{\partial x_i} \frac{\partial \zeta_k}{\partial x_j}$, where $\frac{\partial \bm{\zeta}}{\partial \bm{x}}$ is the inverse Jacobian of the mapping between parametric and physical domains. 

\begin{remark}
    The discrete value of $\phase$ may violate the physical range $\left[-1, 1 \right]$. The resulting value of $\phase$ may yield nonpositive values of density and dynamic viscosity, an issue that may arise especially at high viscosity and density ratios of the immiscible fluid constituents. To avoid this, we follow \cite{khanwale_cpc_2022} and define $\tilde{\phase}^h$ as 
    \begin{equation*}
        \tilde{\phase}^h := \begin{cases}
        \phaseh, & \text{if } |\phaseh| \leq 1. \\
        \operatorname{sign}(\phaseh), & \text{otherwise}.
        \end{cases}        
    \end{equation*}  
    We subsequently compute $\densf = \frac{1}{2} \lbr 1 + \frac{1}{\tilde{\densf}} \rbr + \frac{1}{2} \lbr 1 - \frac{1}{\tilde{\densf}} \rbr \tilde{\phase}^h$ and $\viscf = \viscf \lbr \tilde{\phase}^h \rbr$ in our semi-discrete weak formulation.
\end{remark}

We define the semi-discrete variational multiscale formulation of the coupled fluid-structure interaction problem as follows: find $\bm{U}^{f,h} \in \bm{\mathcal{X}}^{f,h}$, $\bm{U}^{s,h} \in \bm{\mathcal{X}}^{s,h}$, $\bm{U}^{m,h} \in \bm{\mathcal{X}}^{m,h}$ such that $\forall \bm{W}^{f,h} \in \bm{\mathcal{Y}}^{f,h}$, $\forall \bm{W}^{s,h} \in \bm{\mathcal{Y}}^{s,h}$ and $\forall \bm{W}^{m,h} \in \bm{\mathcal{Y}}^{m,h}$, 
\begin{equation}
    B^f_{VMS}(\bm{W}^{f,h}, \bm{U}^{f,h}) + B^{s}(\bm{W}^{s,h}, \bm{U}^{s,h}) + B^{m}(\bm{W}^{m,h}, \bm{U}^{m,h}) = \lbr \xr \wmuh, h_w \lbr \phaseh, \vrh, \vzh; \vrmh, \vzmh \rbr \rbr_{\sfint},
    \label{eqn:weakform_disc}
\end{equation}
\noindent where 

\begin{equation}
\begin{aligned}
    B^f_{VMS}& (\bm{W}^{f,h}, \bm{U}^{f,h}) = B^f (\bm{W}^{f,h}, \bm{U}^{f,h}) + \lbr \xr \frac{\partial \wph}{\partial \xr}, \frac{\taum}{\densf (\tilde{\phase}^h)} R_{mr} \rbr + \lbr \xr \frac{\partial \wph}{\partial \xz}, \frac{\taum}{\densf (\tilde{\phase}^h)} R_{mz} \rbr \\
    & + \lbr \xr \lbr \frac{\partial \wvrh}{\partial \xr} +\frac{\partial \wvzh}{\partial \xz} \rbr + \wvrh, \ \densf(\tilde{\phase}^h) \tauc R_{co} \rbr_{\spatdomf} - \lbr \xr \wvrh, \taum \lbr R_{mr} \frac{\partial \vrh}{\partial \xr} + R_{mz} \frac{\partial \vrh}{\partial \xz} \rbr \rbr_{\spatdomf} \\
    & + \lbr \xr \frac{\partial \wvrh}{\partial \xr}, \taum R_{mr} \lbr \frac{\taum}{\densf (\tilde{\phase}^h)} R_{mr} + \vrh \rbr \rbr_{\spatdomf} + \lbr \xr \frac{\partial \wvrh}{\partial \xz}, \taum R_{mr} \lbr \frac{\taum}{\densf (\tilde{\phase}^h)} R_{mz} + \vzh \rbr \rbr_{\spatdomf} \\
    & - \lbr \xr \wvzh, \taum \lbr R_{mr} \frac{\partial \vzh}{\partial \xr} + R_{mz} \frac{\partial \vzh}{\partial \xz} \rbr \rbr_{\spatdomf} + \lbr \xr \frac{\partial \wvzh}{\partial \xr}, \taum R_{mz} \lbr \frac{\taum}{\densf (\tilde{\phase}^h)} R_{mr} + \vrh \rbr \rbr_{\spatdomf} \\
    & + \lbr \xr \frac{\partial \wvzh}{\partial \xz}, \taum R_{mz} \lbr \frac{\taum}{\densf (\tilde{\phase}^h)} R_{mz} + \vzh \rbr \rbr_{\spatdomf} - \lbr \xr \wch, \frac{\taum}{\densf (\tilde{\phase}^h)} \lbr R_{mr} \frac{\partial \phaseh}{\partial \xr} + R_{mz} \frac{\partial \phaseh}{\partial \xz} \rbr \rbr_{\spatdomf}.
\end{aligned}
\end{equation}

In the present study, we construct finite dimensional trial and weight function spaces using splines through the concept of Isogeometric Analysis (IGA) \cite{cottrell_wiley_2009, hughes_cmame_2005}. We construct the solid displacement fields and their associated weight functions as $\boldsymbol{u}^h (\Xr,\Xz,t) = \sum_{A \in I_s} \boldsymbol{u}^A(t) \widehat{N}_A (\Xr,\Xz)$ and $\boldsymbol{w}^h (\Xr,\Xz,t) = \sum_{A \in I_s} \boldsymbol{w}^A \widehat{N}_A (\Xr,\Xz)$, 
where $A$ is a control variable index, $I_s$ is the index set of the solid control variables and $\widehat{N}_A$ is a spline basis function defined on $\refdom$. We define $\widehat{N}_A$'s as quadratic B-splines that are $\mathscr{C}^1$-continuous everywhere except perpendicular to $\sfintref$, where they are $\mathscr{C}^0$-continuous. To approximate the fluid unknown fields, we push forward $\widehat{N}_A$ to $\spatdom$ such that $N_A (\xr, \xz, t) = \widehat{N}_A \circ \phi^{-1} (\xr, \xz, t) \ \forall A \in I_f$, where $I_f$ is the index set of the fluid control variables. We now construct the fluid pressure field as $\ph (\xr, \xz, t) = \sum_{A \in I_f} p^A(t) N_A (\xr,\xz,t)$ and it's corresponding weight function as $\wph (\xr, \xz, t) = \sum_{A \in I_f} \wp(t) N_A (\xr,\xz,t)$. We approximate the remaining fluid fields and their corresponding weight functions analogously. To approximate the mesh velocity in the spatial configuration, we use $\tilde{\boldsymbol{v}}^h  (\xr, \xz, t) = \sum_{A \in I_f} \frac{\partial \tilde{\boldsymbol{v}}^A (t)}{\partial t} N_A (\xr, \xz, t)$. 

\begin{remark}
To compare our simulation results with experimental measurements of the pressure, we define $p_m = p + \frac{3}{2\sqrt{2}} \ \cahnno \ \ohno^2 | \nabla \phase |^2 $. Under equilibrium conditions, the field $p_m$ varies monotonically across the air-fluid interface. At the interface, $p_m$ is a more faithful representation of the physical pressure than $p$. Far from the air-fluid interface and external fluid boundaries, $\nabla \phase \approx 0$ and $p_m \approx p$. 
\end{remark}

\begin{remark}
   We also implemented an algorithm alternative to Eq.~\eqref{eqn:weakform_disc}. In the alternative algorithm we use $p_m$ as unknown instead of $p$, therefore $p_m$ is spanned by the basis of quadratic splines. We did not observe any significant differences between the two formulations. 
\end{remark}

\subsection{Time discretization and numerical solution strategy}

We use the generalized-$\alpha$ method \cite{chung_1993, jansen_2000}, a fully-implicit time-integration scheme, to time-integrate our semi-discrete fluid-structural equations. To attain second-order accuracy, unconditional stability, and optimal high-frequency damping in our numerical simulations, we select the parameters for the generalized-$\alpha$ method as outlined in \cite{bazilevs_fsirev_2008}. 

To solve the resulting system of nonlinear equations, we use a Newton-Raphson method with a two-stage predictor-multi-corrector approach; see \cite{bazilevs_fsirev_2008} for more details. Our linear solver in each Newton iteration is a Generalized Minimal Residual method (GMRES) \cite{gmres_1986} with a Jacobi preconditioner. We set the relative tolerance of the nonlinear solver to $10^{-4}$. For the mesh motion solver, we set the norm of the change in the solution between each Newton step to $10^{-4}$. We consider that the linear iterative solver has converged when the relative reduction in the preconditioned residual norm is below $10^{-5}$. 

Given the multiscale nature of our FSI problem, we adopt an adaptive time-stepping scheme. Our numerical implementation automatically adjusts the discrete time-step $\Delta t$ based on the number of nonlinear iterations required to attain convergence. Specifically, if our Newton solver converges within three iterations or requires more than five iterations to converge, we accordingly change our time step $\Delta t$ by $5\%$. If our Newton solver diverges, we reduce $\Delta t$ by $20\%$ and restart our time-stepping scheme from the time step immediately preceding divergence. Our estimates show that employing this adaptive time-stepping scheme resulted in computational savings of around $20-30\%$. 

We use a quasi-direct solution strategy \cite{tezduyar_cmame_2006} to solve the coupled equations. This strategy allows us to solve the fluid-solid equations monolithically and the mesh motion equations separately, such that the data for the mesh motion are used from the fluid-structural solver as an input. We develop our code on top of the open source high performance computing libraries PETSc \cite{petsc-web-page} and PetIGA \cite{PETiga_CMAME}.

\section{Results}
\label{sec:results}

\subsection{Description of the simulations and quantities of interest}

\begin{figure}
    \centering
    \includegraphics[width=0.96\linewidth]{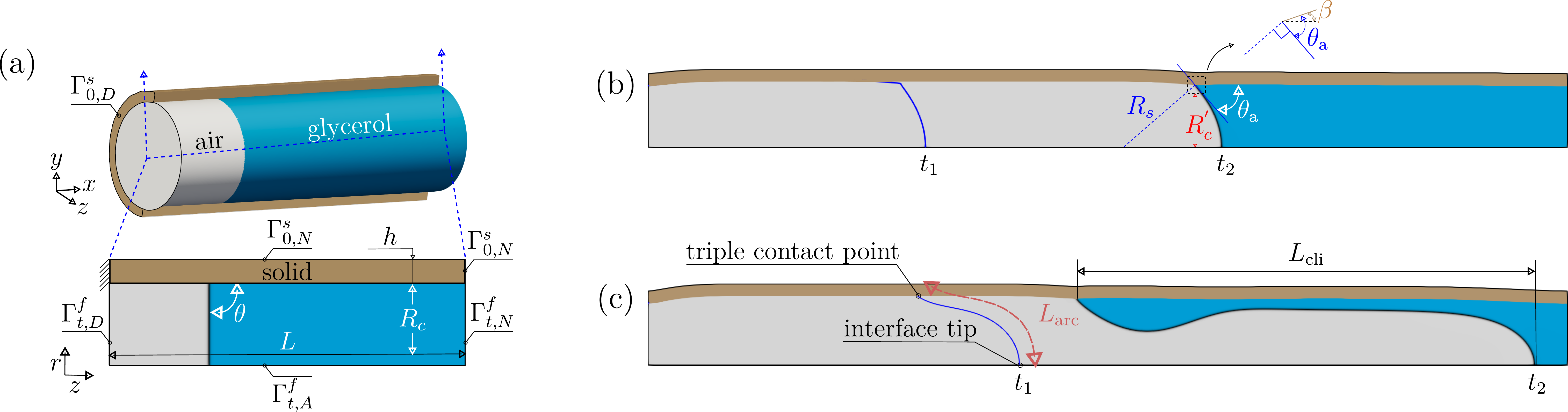}
    \caption{(a) Schematic of the computational domain, initial conditions, domain boundaries (see Eq.~\eqref{eq:strongform_bcs}) and geometrical parameters. Our 2D axisymmetric domain is obtained by slicing the original 3D tube along the center line axis. We only illustrate a portion of the tube in the 3D schematic to show the presence of air and glycerol. (b-c) Air-glycerol interface at two times, $t_1$ and $t_2$. In (b), $R_c^{\prime}$ denotes the radius of the deformed tube and $R_s$ is the radius of a spherical cap fitted to the interface. The zoomed inset in (b) illustrates our definition of the angles $\beta$ and the apparent contact angle $\appcont$. The air-glycerol interface is stable in (b) and unstable in (c). The blue solid line representing the air-glycerol interface at $t_1$ in (b-c) is the level set of $\phase = 0$.}
    \label{fig:schematic}
\end{figure}

We study the displacement by air of a much denser and more viscous liquid in a rigid and a soft capillary tube; see Fig.~\ref{fig:schematic}. We choose the fluid properties and geometrical parameters of the tube to be the same as in the microfluidic experiments performed in a rigid capillary tube \cite{zhao_etal_prl_2018, pahlavan_etal_pnas_2019}. After exploiting the cylindrical symmetry, our computational domain is a two-dimensional rectangular box, with the box length much larger than the box height. We set the reference length scale to $\hat{L}_r = 375 \ \mu$m. The geometry of the capillary tube is defined using three parameters. Unless otherwise specified, we use the following values: initial radius of the capillary tube $R_c=1$, radial thickness of the tube $h = 0.2$, and the length of the tube $L = 15$. We inject air on the left fluid boundary $\fintdir$ at a constant flow rate to displace the liquid. The experiments in \cite{zhao_etal_prl_2018, pahlavan_etal_pnas_2019} used glycerol as the displaced liquid. The viscosity ratio of glycerol to air is $O\lbr 10^5 \rbr$ \cite{zhao_etal_prl_2018}. Modeling such large viscosity ratio requires a very fine mesh to resolve viscosity gradients and small time steps, which can make our computations unaffordable. To address this challenge we use the viscosity ratio $\tilde{\viscf} = 1000$. This is acceptable because the viscous forces in air are negligible with respect to the glycerol's viscous forces and capillary forces. Similar assumptions have been commonly used in prior studies handling gas-liquid systems with negligible gas-phase viscosity \cite{bretherton_jfm_1961, gao_etal_jfm_2019}. Based on the above reasoning, we use a liquid with a dynamic viscosity of $1.4$ $\mathrm{Pa \cdot s}$ and a true density equal to $1260$ $\mathrm{kg/m^3}$. Unless otherwise specified, we take $\tilde{\densf} = 1046.5$ and $\tilde{\viscf} = 1000$. For these fluid parameters, we have $\ohno = 1.14$. 

Unless otherwise stated, we select a solid with properties similar to those of silicone \cite{style_etal_prl_2013} and PDMS gels, i.e., a Young's modulus of $20$ kPa, a mass density of the solid $\denss$ equal to $1000$ $\mathrm{kg/m^3}$ and a Poisson's ratio $\nu = 0.45$. For our selected fluid and solid parameters, we have $\zeta = 0.92$ and $\chi = 8.67 \times 10^{-3}$. We define the wettability of glycerol with an equilibrium contact angle $\theta$; see Fig.~\ref{fig:schematic}. Unless otherwise specified, we select $\pecno = 703$ and $\cahnno = 0.012$. Our selected value of $\cahnno$ is such that the air-glycerol interface is sufficiently thin and the criterion $\cahnno < 4 \pecno^{-0.5} \tilde{\viscf}^{-0.25}$ for the sharp-interface limit is satisfied \cite{yue_etal_jfm_2010}. 

We impose the boundary conditions on the computational domain (see Fig.~\ref{fig:schematic}) as per Eqn.~\eqref{eq:strongform_bcs}. The velocity of injected air is $\bm{v}_\mathrm{in} = \lbr 2 \capno \lbr 1 - r^2 \rbr, 0 \rbr$, where $\capno$ \ is the inlet capillary number. Here, $\capno$ denotes the ratio of viscous forces to the capillary forces. An expression for $\capno$ is given by $\capno = \frac{\eta_\mathrm{l} U_a}{\gamma_\mathrm{la}}$, where $U_a$ is the axial speed of air at $\fintdir$. Our initial conditions in the simulation are as follows: we consider zero fluid velocity, i.e., $\bm{v}_0 = 0$, zero solid displacements, i.e., $\bm{u}_0 = 0$ and we assume that the shape of the air-glycerol interface is initially flat, such that $c_0 = \tanh \frac{\lbr \xz - z_{in} \rbr}{\sqrt{2} \epsilon}$. Here, $z_{in}$ is the axial position of the initial air-glycerol interface and $z_{in} = 1$. {In all of our simulations, we use a uniform mesh with $L \times \lbr R_c + h \rbr \times 10^4$ quadratic elements}. We selected this mesh size after performing simulations with a finer, uniform mesh that had a 2.25-fold increase in quadratic elements. Our results showed less than $1.5 \%$ deviation, confirming the suitability of our chosen mesh size.

In what follows, we study the onset and dynamics of interfacial instability in a  capillary tube. We focus on understanding the impact of $\capno$, $\theta$, and $\zeta$ on interfacial instability. The values of $\capno$ we select in our numerical simulations are within the range of $\capno$ in a rigid capillary tube previously studied using experiments \cite{zhao_etal_prl_2018} and numerical simulations \cite{esmaeilzadeh_etal_pre_2020, gao_etal_jfm_2019}. We study two different values of $\theta$: a wetting case with $\theta = 68^\circ$, which corresponds to drainage, and a non-wetting case with $\theta = 115^\circ$, which corresponds to imbibition. We temporally monitor the shape of the air-glycerol interface using two parameters: the air-glycerol interface arc length $\arclen$ (see Fig.~\ref{fig:schematic}(c) for a definition) and the apparent contact angle of the interface $\appcont$ (see Fig.~\ref{fig:schematic}(b) for a definition). The parameter $\arclen$ quantifies the deformation of the air-glycerol interface. We will also monitor the quantity $\frac{\mathrm{d} \arclen}{\mathrm{d} t}$, which is related to the relative speed between the interface tip and the contact line. When $\frac{\mathrm{d} \arclen}{\mathrm{d} t}$ evolves towards zero, we expect a stable air-glycerol interface. Persistent growth of $\arclen$ is associated with interfacial instability. We define the apparent contact angle $\appcont$ as the angle formed between the tangent to the air-glycerol interface at the triple contact point and the fluid-solid interface. {The axial distance between the triple contact point and the interface tip $\axialcli$ is given by $\axialcli = \frac{R_c^{\prime} \lbr z_{\mathrm{cl}}, t\rbr \big( 1 - \sin (\appcont - \beta) \big)}{\cos (\appcont - \beta)}$, where $\beta$ is the angle made by the fluid-solid interface at the triple contact point with the axial direction and $R_c^{\prime} = R_c^{\prime} \lbr z, t\rbr$ is the dimensionless radius of the deformed tube. In this formula, $R_c^{\prime}$ is computed at the axial position of the triple contact point $z_{\mathrm{cl}}$.} This formula was used in \cite{fermigier_jcis_199} for a rigid capillary tube with $\beta = 0$, and later in \cite{suo_wrr_2024} for an inclined capillary tube with a constant $\beta$. This formula is derived by approximating the air-glycerol interface as a spherical cap with radius $R_s = R_c^{\prime}\lbr \cos \lbr \appcont - \beta\rbr \rbr^{-1}$; see Fig.~\ref{fig:schematic}(b) for more details. This spherical cap passes through the interface tip and the contact line, with the tangent to the spherical cap at the interface tip being perpendicular to the axial direction. Knowing $\axialcli, R_c^{\prime}$ and $\beta$ from our simulation data, we compute $\appcont$ as, 
\begin{equation}
    \appcont = \sin^{-1} \lbr \frac{1}{\sqrt{1 + \lbr \sfrac{\axialcli}{R_c^{\prime}} \rbr^2}}\rbr - \tan^{-1} \lbr \sfrac{\axialcli}{R_c^{\prime}} \rbr + \beta.
\end{equation}
The spherical cap approximation can be used when the air-glycerol interface is stable and $\appcont > 0$. When $\appcont < 0$, the air-glycerol interface can no longer be approximated with a spherical cap. Following \cite{zhao_etal_prl_2018}, the time when $\appcont = 0$ marks the onset of interfacial instability.

\subsection{Forced imbibition and drainage in a rigid capillary tube}

We study the onset of interfacial instability in a rigid capillary tube for imbibition and drainage. This study serves two purposes: a) first, it provides a baseline for later comparing the results of interfacial instability in a soft capillary tube; b) second, it validates our computational approach for the two-fluid problem. For the simulations in this section, we do not solve the equations of solid dynamics or mesh motion.

\subsubsection{Effect of inlet capillary number on interface deformation and interfacial instability}

We study two key aspects here: (i) the effect of $\capno$ on the interface deformation, and (ii) the minimum $\capno$ required to trigger an interfacial instability, which we define as the critical inlet capillary number $\capnocr$. 

\begin{figure}[t] 
    \centering
    \begin{overpic}[width=0.9\linewidth]{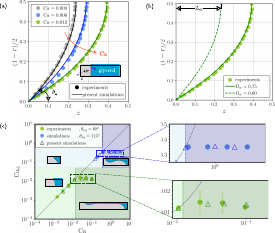}
        \put(26.0,37){{\tiny \cite{zhao_etal_prl_2018}}}
        \put(25.5,34.5){{\tiny \cite{esmaeilzadeh_etal_pre_2020}}}
        \put(39.5,52.5){{\tiny \cite{zhao_etal_prl_2018}}}
        \put(95.,55.5){{\tiny \cite{zhao_etal_prl_2018}}}        
    \end{overpic}
    \caption{Comparison of our simulations with  experiments \cite{zhao_etal_prl_2018} and previous numerical simulations \cite{esmaeilzadeh_etal_pre_2020} in a rigid capillary tube. (a) Steady state air-glycerol interface in the wetting capillary tube ($\theta = 68^\circ$) for different $\capno$. 
    (b) Steady state air-glycerol interface in the wetting capillary tube ($\theta = 68^\circ$) for two different values of $\dw$ when $\capno = 0.012$. (c) Contact line capillary number $\capnocl$ as a function of $\capno$ in wetting ($\theta = 68^\circ$) and non-wetting ($\theta = 115^\circ$) capillary tubes. 
    The green and blue vertical dotted lines represent the critical inlet capillary numbers $\capnocr$. The insets in the left subfigure show  sketches of the stable glycerol-air interface regime when $\capno <  \capnocr$ and unstable air-glycerol interface regime when $\capno > \capnocr$. The zoomed insets on the right subfigure show a close up of the comparison between our simulation results and the results from the literature \cite{zhao_etal_prl_2018, esmaeilzadeh_etal_pre_2020}.}
    \label{fig:validation_1}
\end{figure}
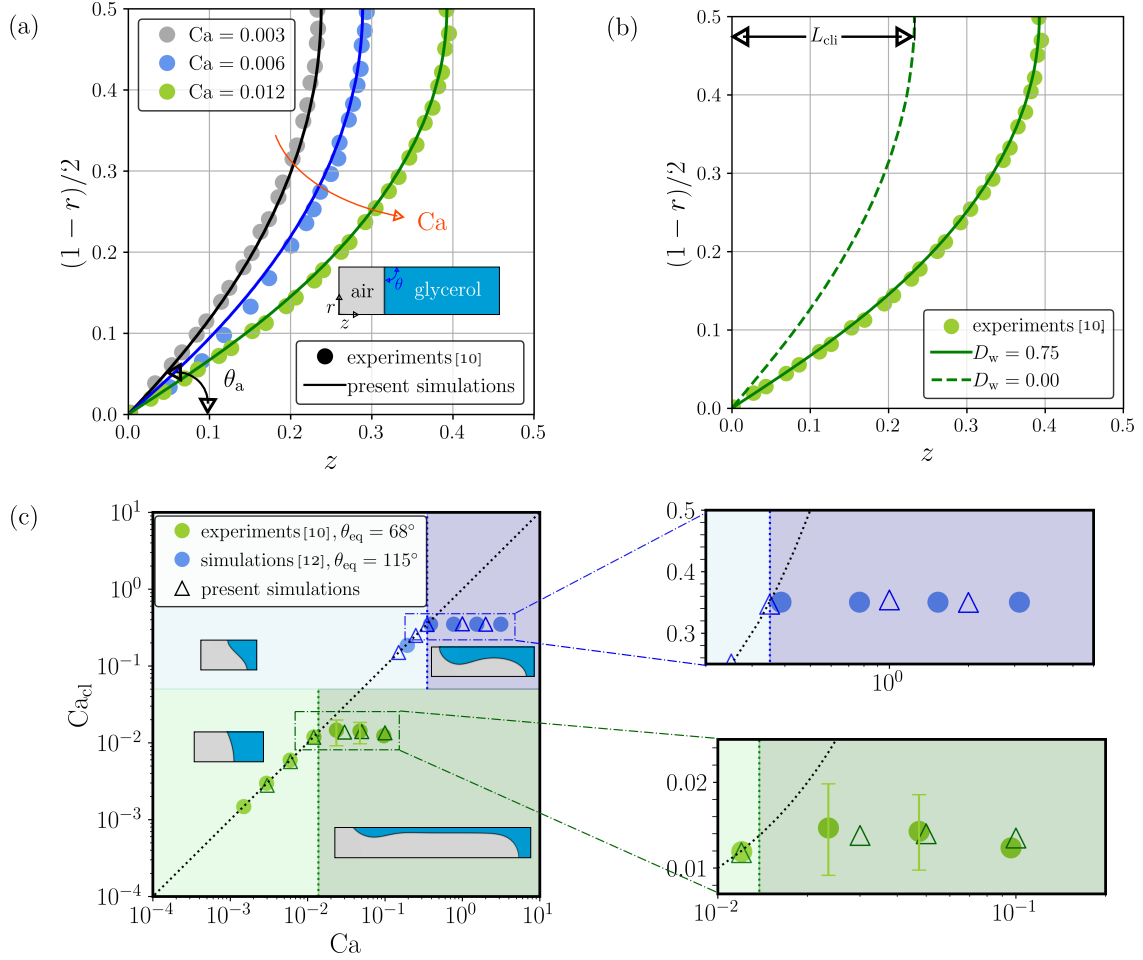

We begin by first studying cases with a low $\capno$ for which the interface reaches a stationary state. As shown in Fig.~\ref{fig:validation_1}(a), as $\capno$ increases, the air-glycerol interface bends closer to the walls of the capillary tube. We also use these computations to illustrate the importance of the dynamic contact angle boundary condition. To obtain an agreement between the air-glycerol interface profiles obtained from our simulations and experiments \cite{zhao_etal_prl_2018}, we select $\dw = 0.75$ for $\theta = 68^\circ$ for all values of $\capno$. We observe that when $\dw = 0$ is selected for $\capno = 0.012$, the air-glycerol interface profile, and consequently $\axialcli$, is significantly under-predicted; see Fig.~\ref{fig:validation_1}(b). This under-prediction of $\axialcli$ leads to an over-prediction of $\capnocr$ by $\approx 150\%$---this indicates that using a dynamic wettability condition is critical to match experimental data. Although one can find in the literature dynamic contact angle laws more sophisticated than Eq.~\eqref{eqn:dynamic_contactangle_law} (e.g., using the traditional Cox-Voinov’s relationship as done in \cite{esmaeilzadeh_etal_pre_2020} or modifying the free energy density as done in \cite{qiu_thesis_2024}), our data indicate that Eq.~\eqref{eqn:dynamic_contactangle_law} is robust and accurate enough for computing the interface deformation and $\capnocr$. Importantly, while accurate predictions using Eq.~\eqref{eqn:dynamic_contactangle_law} require the use of a value of $\dw$ that depends on $\theta$, this value does not depend on any other parameter of the problem.

To study the onset of interfacial instability, we plot the contact line capillary number $\capnocl$ against $\capno$ for wetting and non-wetting rigid capillary tubes in Fig.~\ref{fig:validation_1}(c). Here, $\capnocl = \frac{\eta_\mathrm{l} U_\mathrm{cl}}{\gamma_\mathrm{la}}$, where $U_\mathrm{cl}$ is the speed of the contact line. We define the contact line as the intersection of the level set $\phase = 0$ with the walls of the capillary tube. In our computations, $\capnocl$ is calculated as the constant slope of the contact line position data over time; see \cite{esmaeilzadeh_etal_pre_2020} for more details. For the non-wetting case, we use $D_w = 5 \times 10^{-5}$ in the simulations. Fig.~\ref{fig:validation_1}(c) shows good agreement between our numerical results and previous experimental and simulation data \cite{zhao_etal_prl_2018, esmaeilzadeh_etal_pre_2020} for a wide range of $\capno$. Our simulation results in Fig.~\ref{fig:validation_1}(c) support four findings presented in \cite{zhao_etal_prl_2018, esmaeilzadeh_etal_pre_2020}. First, there exists a critical capillary number $\capnocr$, such that, for $\capno < \capnocr$, the interface remains stable and for $\capno > \capnocr$, the interface becomes unstable. The values of $\capnocr$ estimated from our numerical simulations are $\capnocr = 0.0138$ for $\theta = 68^\circ$ and $\capnocr = 0.35$ for $\theta = 115^\circ$, both of which are within $2\%$ the values reported in \cite{zhao_etal_prl_2018, esmaeilzadeh_etal_pre_2020}. Second, when $\capno < \capnocr$, the speeds of the contact line and injected air at the inlet are equal, i.e., $\capnocl = \capno$. Third, when $\capno > \capnocr$, $\capnocl$ becomes independent of $\capno$ and $\capnocl = \capnocr$. Fourth, $\capnocr$ depends only on the wettability of the tube and $\capnocr$ increases with $\theta$. Zoomed insets in Fig.~\ref{fig:validation_1}(c) also show that our simulation results for $\capnocl$ deviate by less than $2\%$ from previously reported data \cite{zhao_etal_prl_2018, esmaeilzadeh_etal_pre_2020}.

\subsubsection{Comparison of the computational prediction of the film thickness with theory and experiments}

When the interface becomes unstable, a thin film of glycerol is entrained along the walls of the capillary tube. If the radius of the tube is sufficiently small for gravity to be negligible, we expect that this film will have uniform thickness ($h_f$) between the front and rear menisci. We compute $h_f$ for a range of interface tip capillary numbers $\capnotip$ for cases with $\capno > \capnocr$. Here, $\capnotip = \frac{\eta_\mathrm{l} U_\mathrm{tip}}{\gamma_\mathrm{la}}$, where $U_\mathrm{tip}$ is the speed of the interface tip. We will compare the results of our numerical simulations with the expressions
\begin{alignat}{2}
    &h_f = 1.34(\capnotip)^{2/3} \label{eqn:filmthickness_exprs_1}, \\
    &h_f =     \frac{1.34(\capnotip)^{2/3}}{1 + 2.5 \times 1.34(\capnotip)^{2/3}}.
    \label{eqn:filmthickness_exprs_2}
\end{alignat}
Eq.$~\eqref{eqn:filmthickness_exprs_1}$ was determined by Bretherton \cite{bretherton_jfm_1961} and is valid only for thin films with negligible inertia, characterized by $\capnotip < 5 \times 10^{-3}$. Eq.$~\eqref{eqn:filmthickness_exprs_2}$ is known as {\it Taylor's law} and is valid for films with larger inertia. Eq.$~\eqref{eqn:filmthickness_exprs_2}$ is usually assumed valid for $\capnotip < 2$. Eq.$~\eqref{eqn:filmthickness_exprs_2}$ was proposed in \cite{aussillous_pof_2000} by fitting Taylor's experimental data \cite{taylor_jfm_1961}. The factor $2.5$ in the denominator of Eq.$~\eqref{eqn:filmthickness_exprs_2}$ is empirical. Given $\capno$, we compute $h_f$ and $\capnotip$ from Eq.~\eqref{eqn:filmthickness_exprs_1} by solving it in conjuction with $\capno = (1 - h_f)^2 \capnotip$, which follows from mass conservation. We solve this nonlinear equation using a modified Powell hybrid numerical method \cite{more_report_1980}, which is a combination of the Newton's method and the secant method. We follow the same procedure to compute $h_f$ and $\capnotip$ from Eq.~\eqref{eqn:filmthickness_exprs_2} given a $\capno$. 

\begin{figure} 
    \centering
    \begin{overpic}[width=0.8\linewidth]{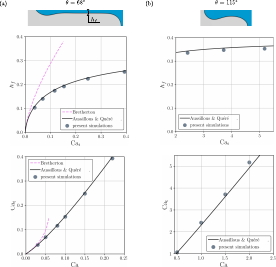}
        \put(34.5,55.3){{\tiny \cite{bretherton_jfm_1961}}}
        \put(41.,52.9){{\tiny \cite{aussillous_pof_2000}}} 
        \put(23.3,37){{\tiny \cite{bretherton_jfm_1961}}}    
        \put(29.7,34.5){{\tiny \cite{aussillous_pof_2000}}}
        \put(83.7,53.2){{\tiny \cite{aussillous_pof_2000}}}
        \put(94.5,9.3){{\tiny \cite{aussillous_pof_2000}}} 
    \end{overpic}
    \caption{Variation of the thickness of entrained film $h_f$ with the interface tip capillary number $\capnotip$ (top) and dependence of $\capnotip$ on the inlet capillary number $\capno$ (bottom) for a rigid tube. Panel (a) shows the wetting case ($\theta = 68^\circ$) and panel (b) shows the non-wetting case ($\theta = 115^\circ$).}
    \label{fig:validation_2}
\end{figure}
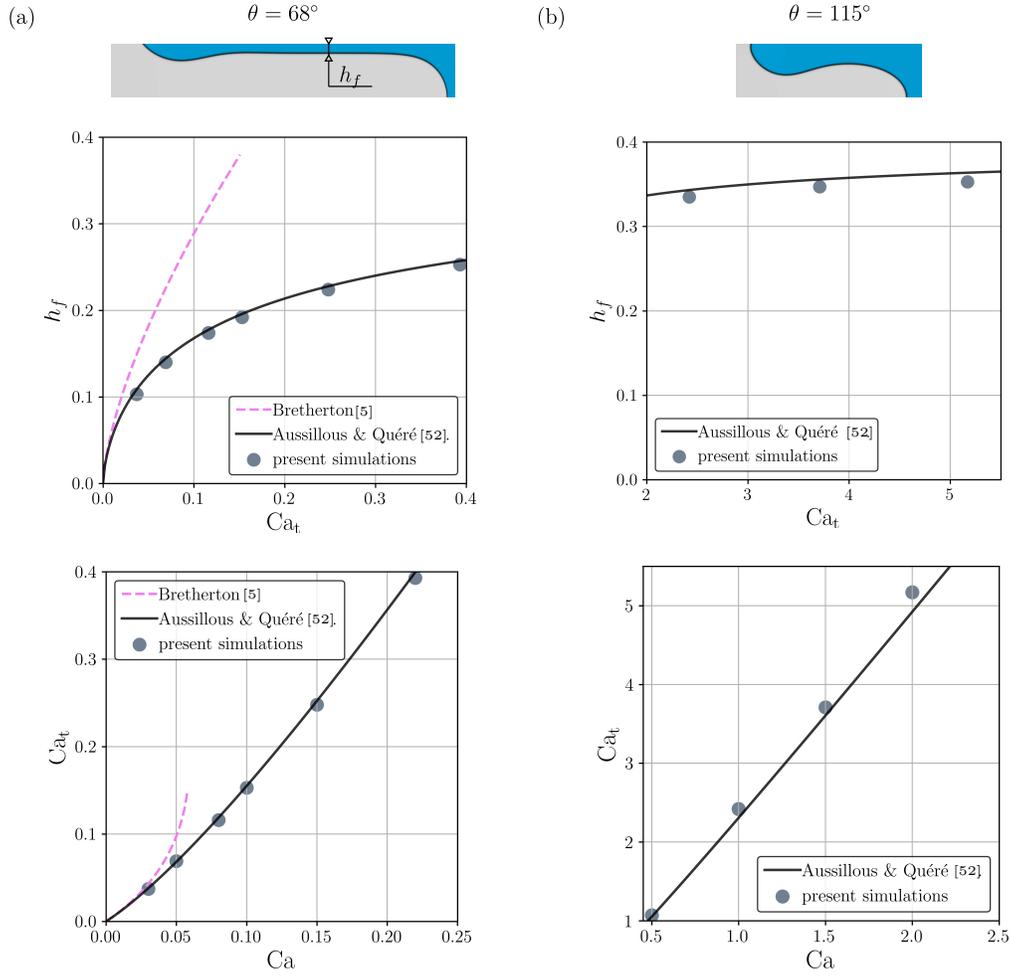

In our numerical simulations, we estimate $h_f$ by computing the distance between the level set $\phase = 0$ and the tube wall (see the schematic in Fig.~\ref{fig:validation_2}) once the subsequent change in its value is less than $5\%$; see \cite{esmaeilzadeh_etal_pre_2020} for more details. We define the interface tip as the intersection of the level set $\phase = 0$ with the symmetry axis of the tube; see Fig.~\ref{fig:schematic}(c). We compute $\capnotip$ as the constant slope of the interface tip position data over time. 

Fig.~\ref{fig:validation_2}(a) shows the results for the wetting case $\theta=68^\circ$. The top plot shows that our simulations agree well with \cite{aussillous_pof_2000} for all values of $\capnotip$. As expected, Bretherton's law \cite{bretherton_jfm_1961} works well for very low $\capno$, but becomes very inaccurate for large $\capno$. The bottom panel of Fig.~\ref{fig:validation_2}(a) shows a similar trend in the prediction of $\capnotip$ for a range of values of $\capno$. Fig.~\ref{fig:validation_2}(b) presents the results for the non-wetting case. Because reaching the instability in the non-wetting case requires much larger values of $\capno$ we do not plot the predictions of Bretherton's law, which are very inaccurate. However, the predcitions from Taylor's law \cite{aussillous_pof_2000} agree well with our simulations even for cases with $\capnotip>2$. Although the agreement is not as good as it was for the wetting case (lower values of $\capno$), the results indicate that Taylor's law leads to an {error smaller than $0.7$}\% for $\capnotip<5$. 

\subsection{Forced drainage in a soft capillary tube}
Here, we show that the tube's compliance delays the interfacial instability in a forced drainage process. We perform simulations in a soft wetting capillary tube with $\theta = 68^\circ$. For the simulation results reported in this subsection, we use $\dw = 0.75$ because this is the value that leads to simulations that best match experimental data in a rigid tube. Here, we study the effects of two dimensionless parameters ($\capno$ and $\zeta$) on the stability of the air-glycerol interface. To determine the interface stability and the onset time of interfacial instability, we plot the time evolutions of $\arclen$ and $\appcont$, respectively, because they are reliable indicators of the air-glycerol interface deformation. {We also show that the comparison of $\capnocl$ and $\capno$ used for a rigid tube (see Fig.~\ref{fig:validation_1}) is not a good indicator of the interface stability in a soft tube.}

\subsubsection{Effect of inlet capillary number on the air-glycerol interface stability}

We first study the stability of the air–glycerol interface as a function of $\capno$. {Here, the tube's compliance is given by $\zeta = 0.92, \ \chi = 8.67 \times 10^{-3}$ and is the same for all values of $\capno$. For comparison purposes, we will also perform simulations for a rigid tube, i.e., $\zeta = \infty$ and $\chi = 0$.} We use four different values of $\capno$, namely, 0.012, 0.015, 0.04, and 0.4. The case of $\capno = 0.012$ corresponds to a stable interface in soft and rigid tubes. The cases defined by $\capno = 0.015$, 0.04, and 0.4 correspond to an unstable interface in the rigid and the compliant tube--- however, the instability occurs at a later time in the compliant tube. 

\begin{figure}
    \centering
    \includegraphics[width=0.99\linewidth]{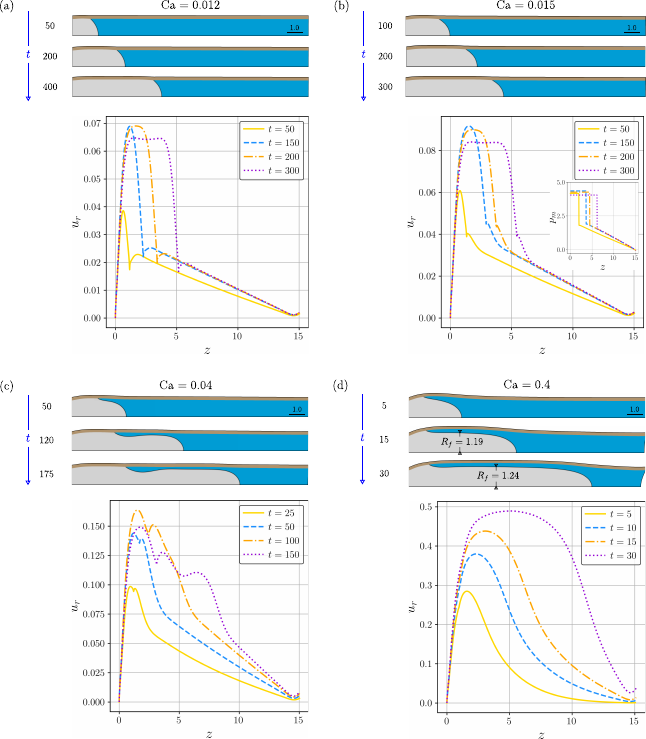}
    \caption{Time evolution of air-glycerol interface profiles and radial solid displacements at the fluid-solid interface in a soft wetting capillary tube ($\theta = 68^\circ$) for different values of inlet capillary number ($\capno$). Air is gray, glycerol is blue and the solid is brown. The inset in (b) shows the time variation of the pressure $p_m$ along the symmetry axis. In (d), the radius of the finger $R_f$ at its midpoint is specified at two different times.}
    \label{fig:softwet_capno}
\end{figure}

Fig.~\ref{fig:softwet_capno} shows snapshots of the solution for different values of $\capno$. At low $\capno$, capillary forces dominate over viscous forces, stabilizing the air-glycerol interface and preventing any significant air-glycerol interface deformation. This behavior is exemplified by panels (a) and (b), where $\capno = 0.012$ and 0.015, respectively. As $\capno$ increases, viscous forces become stronger relative to the capillary forces, eventually triggering an interfacial instability. This instability begins when a finger propagates through the center of the capillary tube, entraining a thin film of glycerol along the walls of the tube; see panels (c-d). This film entrainment behavior has been previously reported in a rigid capillary tube for unstable cases \cite{zhao_etal_prl_2018, esmaeilzadeh_etal_pre_2020}. However, in contrast to a rigid tube, the radius of the finger here varies in time and space.

We can get further insight in the process by analyzing the deformation of the tube examining the radial solid displacement at the fluid-solid interface. The tube expands in a spatially heterogeneous manner that is characterized by a peak and a sharp local minimum at the triple contact point. The peak displacement increases with $\capno$, as shown in Fig.~\ref{fig:softwet_capno}, and is driven by the pressure forces from the advancing air phase. Because we are considering the wetting case, the curvature of the interface is such that the pressure grows when we cross the interface going from glycerol to air. This can be seen in the inset of panel (b), which shows the fluid pressure at the symmetry axis. The local minimum in the radial solid displacement is formed when capillary forces pull the solid radially towards the symmetry axis; see panels (a-c). This local minimum in radial solid displacement is similar to the wetting ridges observed in elastocapillarity problems \cite{sauer_mms_2016, style_etal_prl_2013}. At very high $\capno = 0.4$, this local minimum is not noticeable because the pressure-driven deformation of the tube dominates the overall solid deformation.

\begin{figure*}
    \centering
    \includegraphics[width=0.85\linewidth]{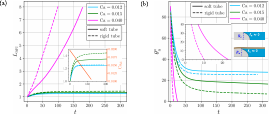}
    \caption{{Air-glycerol interfacial instability metrics in a wetting capillary tube ($\theta = 68^\circ$) for three different values of inlet capillary number ($\capno$). (a) Temporal variation of interface arc length $\arclen$. The inset shows also the time evolution of $\capnocl$ for $\capno = 0.04$. (b) Time evolution of apparent contact angle $\appcont$. The insets show the snapshots of the solution in both rigid and soft tubes for $\capno = 0.04$ near the onset of instability. In the insets, $R_s$ denotes the radius of the spherical cap, which at that specific time instant, is nearly perpendicular to the fluid-solid interface profile at the triple contact point. In (a-b), the solid lines denote the results in a soft tube and the dashed lines denote the results in a rigid tube. The air-glycerol interface is unstable for $\capno = 0.015$ and $0.040$, but stable for $\capno = 0.012$.}}
    \label{fig:stretch_appangle_wet}
\end{figure*}

{To further study the interfacial instability, we track the time evolution of $L_\mathrm{arc}$ and $\appcont$ for different values of $\capno$; see Fig.~\ref{fig:stretch_appangle_wet}. For all values of $\capno$, the air-glycerol interface stretches from its initial length of $1$, as shown in panel (a). When the air-glycerol interface is stable at low $\capno = 0.012$, the value of $L_\mathrm{arc}$ in both the soft and rigid tubes does not change significantly for $t \gtrsim 60$, indicating that the interface tip and the contact line eventually move at the same speed. The difference in the magnitude of $\arclen$ between the soft tube and the rigid tube (see the inset in panel (a)) is very small. This is because the tube’s deformation at a low $\capno$ is insufficient to significantly affect the interface deformation. For $\capno = 0.015$, we observe a slow increase of $L_\mathrm{arc}$ over time in both the soft and rigid tubes; see the zoomed inset in panel (a). Although we did not observe an instability in either of the simulations, we believe that this might be because the tube's length of the computational domain was not large enough for the instability to develop. When $\capno$ is large, i.e., $\capno = 0.04$, the system rapidly transitions to an unstable regime, where $L_\mathrm{arc}$ seems to grow superlinearly with time. In this case, $\frac{{\rm d}^2 \arclen}{{\rm d} t^2} > 0$ at late times, indicating that the relative speed between the interface tip and contact line increases with time. For a rigid tube at $\capno = 0.04$, $\arclen$ grows much faster than in a soft tube, which indicates that the relative speed between the interface tip and the contact line increases much more rapidly in the rigid tube. Panel (a) shows that, at any given time $t$, the deformation of the air-glycerol interface is always greater in a rigid tube than in a soft tube for the same $\capno$.} 

{To analyze if the relationship between $\capnocl$ and $\capno$ is a good indicator of the interface stability here, we plot the time evolution of $\capnocl$ in a soft tube for $\capno = 0.04$; see the inset in panel (a). We first track the axial position data of the contact line $z_\mathrm{cl}$ and fit a quadratic B-spline to these data. Differentiating this spline with respect to time gives us $\capnocl$. As shown in the inset of panel (a), $\capnocl$ is not constant---hence, we expect the plot of $\arclen$ to be a better indicator of the interface stability.}

When $\capno = 0.012$ in a soft tube, panel (b) shows a small decline ($\approx 1\%$) in $\appcont$ for $t > 100$. Although the air-glycerol interface is stable, this slow reduction in $\appcont$ is caused by the decrease in the radial displacement of the solid at the triple contact point; see panel (a) in Fig.~\ref{fig:softwet_capno}. A temporal decrease in the radial solid displacement for a constant $\arclen$ at steady state necessarily reduces $\appcont$. However, we observe that $\appcont$ remains essentially constant for $t > 100$ in the rigid tube at $\capno = 0.012$. The transition to an instability in both soft and rigid tubes for $\capno = 0.015$ is very gradual. The late onset of interfacial instability at $\capno = 0.015$ in a rigid tube is justified because this case is very close to $\capnocr$; see Fig.~\ref{fig:validation_1}. To observe the onset of interfacial instability in this case, we would need to use a tube that is much longer than the current tube used in our simulations. The transition to an instability in a soft tube, however, occurs much faster at $\capno = 0.04$ compared to $\capno = 0.015$---at $t \approx 23$, $\theta_\mathrm{a} \approx 0^\circ$. The instability in the rigid tube at $\capno = 0.04$ occurs much earlier (at $t \approx 10$) than in a soft tube. We can conclude from panel (b) that, at any given time $t$, the value of $\appcont$ is always smaller in a rigid tube than in a soft tube for the same $\capno$.

\subsubsection{Effect of elastocapillary number on the air-glycerol interface stability}

\begin{figure}[t] 
    \centering
    \begin{overpic}[width=0.95\linewidth]{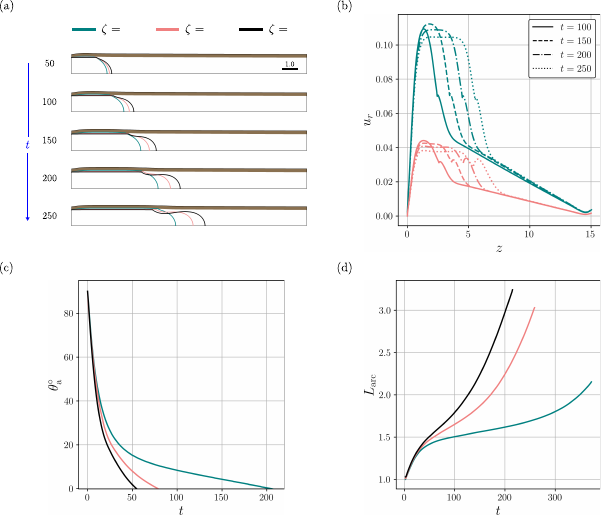}
        \put(20.5,80.3){{\tiny 0.92}}
        \put(35,80.3){{\tiny 2.31}}
        \put(48,80.3){{\tiny $\infty$}}
    \end{overpic}
    \caption{Effect of the tube's compliance on the onset of interfacial instability in a wetting capillary tube ($\theta = 68^\circ$) for inlet capillary number $\capno = 0.02$. a) Time evolution of the air-glycerol interface for different values of $\zeta$. b) Spatio-temporal variation of the radial solid displacement at the fluid-solid interface. Time evolution of the c) apparent contact angle $\appcont$ and the d) air-glycerol interface arc length $\arclen$. The interface is unstable for all cases of $\zeta$ shown in this figure.}
    \label{fig:deformability_wet}
\end{figure}

{Here, we investigate the effect of elastocapillary number on interfacial instability. We compare three different cases: $\lbr \zeta, \chi \rbr = \lbr 0.92, 8.67 \times 10^{-3} \rbr, \lbr 2.31, 3.47 \times 10^{-3} \rbr$ and $\lbr \infty, 0 \rbr$, the last corresponding to a rigid tube. All three cases are computed at $\capno = 0.02$; see Fig.~\ref{fig:deformability_wet}}. As shown in panel (a), the air-glycerol interface propagates most slowly when $\zeta = 0.92$, faster when $\zeta = 2.31$, and fastest in the rigid tube. The difference in the air-glycerol interface speed for different values of $\zeta$ can be explained as follows: As $\zeta$ decreases, the tube’s expansion increases, which by mass conservation, leads to a reduction of the local fluid velocity. Here, we quantify the tube’s expansion by the radial solid displacement at the fluid-solid interface, which is $\approx 3$ times larger for $\zeta = 0.92$ compared to $\zeta = 2.31$; see panel (b). The data in panel (c) show that the onset of the instability ($\theta_a=0$) occurs at $t = 55$ for the rigid tube, $t = 80$ for $\zeta = 0.92$, and $t = 205$ for $\zeta = 2.31$. As $\zeta$ decreases, the tube’s expansion weakens the viscous forces relative to the capillary forces, making the system less prone to an interfacial instability. This observation is supported by the time evolution of $\arclen$ in panel (d), which shows that $\arclen$ and $\frac{\mathrm{d} L_\mathrm{arc}}{\mathrm{d} t}$ grow as $\zeta$ increases, indicating that stiffer tubes are more susceptible to an interfacial instability.

\subsection{Forced imbibition in a soft capillary tube}

In this subsection, we show that the tube's compliance can either delay or suppress the interfacial instability in a forced imbibition process. We perform simulations in a soft non-wetting capillary tube with $\theta = 115^\circ$. For the simulation results reported here, we use $\dw = 5 \times 10^{-5}$ because it is the value for which our simulations best match the data in the literature for a rigid non-wetting tube. Unless otherwise specified, we select $\nu = 0.25$ and $\ohno = 3.60$ here. We have selected a lower Poisson's ratio and a higher Ohnesorge number than in the forced drainage system because the effect of the radial tube's deformation on the interface tip position relative to the contact line becomes more pronounced and easier to analyze. We study the effects of $\capno$ and $\zeta$ on the stability of the air-glycerol interface.

\subsubsection{Effect of inlet capillary number on the air-glycerol interface stability}

{First, we study the influence of $\capno$ on the stability of the air-glycerol interface using $\zeta = 0.35$, $\chi = 2.31 \times 10^{-2}$ and $L = 8$ for all values of $\capno$}. We also compare our solutions with those in a rigid tube, i.e., $\zeta = \infty$. We consider four different values of $\capno$, namely, 0.01, 0.05, 0.1, and 0.5. The cases of $\capno = 0.01, 0.05$ and $0.1$ correspond to a stable interface in both soft and rigid tubes. However, for $\capno = 0.5$ the interface is unstable in the rigid tube but stable in the compliant tube.

Although we find that the tube’s compliance also delays or suppresses interfacial instability, its effect is significantly more complex in the imbibition case than in the drainage case. In the drainage case, the system becomes more stable because the tube expands leading to smaller fluid velocity. In the imbibition case at small $\capno$, we have two counteracting mechanisms: a) the air-glycerol interface curvature is opposite to that in the drainage case. As a consequence the interface tip lags behind the contact line, which makes the system even more stable; b) the air-glycerol interface curvature in the imbibition case leads to a pressure drop when we cross the interface going from glycerol to air. This causes a reduction in the cross section of the tube that makes the system more prone to instability. The relative strength of these two mechanisms determines the overall stability of the air-glycerol interface. 

\begin{figure}
    \centering
    \includegraphics[width=0.9\linewidth]{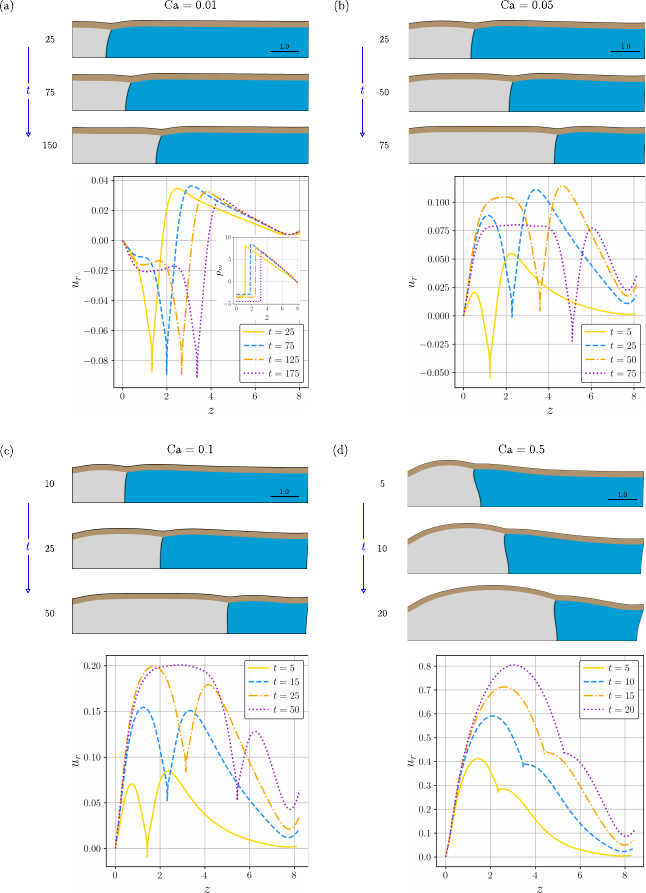}
    \caption{Time evolution of the air-glycerol interface and radial solid displacements at the fluid-solid interface in a soft non-wetting capillary tube ($\theta = 115^\circ$) for different values of inlet capillary number $\capno$. The inset in (a) shows the time evolution of the pressure $p_m$ along the symmetry axis.}
    \label{fig:softnonwet_capno}
\end{figure}

{Fig.~\ref{fig:softnonwet_capno} shows snapshots of our solution for different values of $\capno$. At low $\capno$, capillary forces remain stronger than the viscous forces. Due to the non-wetting nature of glycerol, the air-glycerol interface adopts a curvature where the interface tip lags behind the contact line; see panel (a) for $\capno = 0.01$. The interface curvature is now opposite to that observed in the forced drainage case. As shown in panel (a), the tube contracts, thereby increasing the viscous forces but the stabilizing effect of the interface curvature prevails, and the interface remains stable. At intermediate $\capno$ (0.05 and 0.1), the air-glycerol interface curvature flattens and the axial position of the interface tip slowly approaches the contact line. For these values of $\capno$, the tube expands, thereby weakening the viscous forces. The stabilizing effect of weakened viscous forces leads to interface stability. At high $\capno = 0.5$, the air-glycerol interface curvature reverses such that the interface tip is now ahead of the contact line—this is similar to the forced drainage case; see panel (d). The tube expands in this case, but the resulting weakening of the viscous forces is still strong enough to outweigh the destabilizing effect of the interface curvature—the air-glycerol interface thus remains stable. We expect the air-glycerol interface to trigger an instability with subsequent film entrainment behavior at very high $\capno$, once the destabilizing effect of the interface curvature surpasses the stabilizing effect of the weakened viscous forces.}

We now analyze the radial solid deformation at the fluid-solid interface shown in Fig.~\ref{fig:softnonwet_capno}. We observe a local minimum in the radial solid displacement at the triple contact point, but it is much stronger than in the forced drainage case. Two aspects are notably different as compared with with the forced drainage system. First, at low $\capno = 0.01$, the curvature of the air-glycerol interface causes the pressure to drop across the interface from glycerol to air. This can be seen in the inset of panel (a), where we show the fluid pressure along the symmetry axis. Second, the tube locally contracts from the inlet upto the triple contact point at low $\capno$---this is caused by the negative fluid pressure (relative to the ambient, which we have taken to be zero) in the advancing air phase; see the inset of panel (a). 

\begin{figure*}
    \centering
    \includegraphics[width=0.85\linewidth]{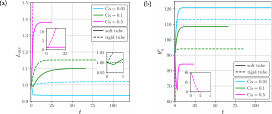}
    \caption{{Time evolution of interfacial instability metrics in a non-wetting capillary tube ($\theta = 115^\circ$) for three different values of inlet capillary number ($\capno$). (a) Temporal variation of interface arc length $\arclen$. The insets shows the evolution of $\arclen$ for $\capno = 0.1$ and $\capno = 0.5$ in a certain time period. (b) Temporal evolution of apparent contact angle $\appcont$. The inset shows the evolution of $\appcont$ for $\capno = 0.5$ in a rigid tube. The air-glycerol interface is stable for $\capno = 0.01$ and $0.05$ in both rigid and soft tubes. For $\capno = 0.5$, the interface is stable in a soft tube but unstable in a rigid tube.}}
    \label{fig:stretch_appangle_nonwet}
\end{figure*}

{To further understand the interface stability, we plot the time evolution of $\arclen$ and $\appcont$ for three different values of $\capno$; see Fig.~\ref{fig:stretch_appangle_nonwet}. Unlike in the forced drainage case, where the air-glycerol interface always stretches due to the tube’s expansion, in the imbibition case the interface may stretch or contract. At $\capno = 0.01$,  $\arclen$ decreases by $\sim 6 \%$ from its initial value of $1$ in the soft tube; see panel (a). This behavior is due to the tube’s contraction and a subsequent change in the interface curvature, as shown in panel (a) of Fig.~\ref{fig:softnonwet_capno}. For a rigid tube at $\capno = 0.01$, the  interface stretches by $\sim 3\%$ in the computed time interval. At an intermediate $\capno = 0.1$, the tube contracts briefly up to $t \approx 3$ slightly reducing $\arclen$ from its initial value; see the inset in panel (a). Over time, however, the tube  expands and stretches the air-glycerol interface by up to $10 \%$. In the rigid tube at the same $\capno$, the  interface stretches up to $15 \%$. When $\capno$ is large in the soft tube, i.e., $\capno = 0.5$, the interface still remains stable and stretches by up to $38\%$. However, in the rigid tube at the same $\capno$, $\arclen$ grows with time and the system rapidly transitions to an unstable regime; see the inset in panel (a). For $\capno = 0.1$ and $\capno = 0.5$, we have plotted $\arclen$ and $\appcont$ only up to the time at which the air-glycerol interface has advanced to an axial position of $\approx 0.8 L$. That is why $\arclen$ curves in panel (a) do not extend to $t > 100$. We did not plot $\arclen$ at later times because its value may be affected by the outlet boudary conditions. As previously observed in a forced drainage case, panel (a) indicates that in the imbibition case $\arclen$ is always higher for a rigid tube than in a soft tube for the same $\capno$. Due to the non-wetting nature of glycerol in forced imbibition, $\appcont$ always exceeds $90^\circ$ for low and intermediate values of $\capno$; see panel (b). However, for high values of $\capno$, $\appcont$ always remains lower than $90^\circ$---this trend is the same as in a forced drainage case. For $\capno = 0.01$ and $0.1$, $\appcont$ remains approximately constant for $t \gtrsim 10$ in soft and rigid tubes. At $\capno = 0.1$ in the rigid tube, $\appcont$ deviates very little from its initial value of $90^\circ$. In the soft tube, however, at the same $\capno$, the local tube deformation near the triple point leads to $\appcont \approx 109^\circ$. At high $\capno = 0.5$ in a soft tube, $\appcont$ drops sharply until $t \approx 5$ and then remains nearly constant for $t \gtrsim 10$. This brief decline and subsequent rise in $\appcont$ is because, at early times, the interface tip moves rapidly ahead of the contact line. Over time, as the tube continues to expand and the local fluid velocity reduces, the advancement of the interface tip relative to the contact line decreases, eventually stabilizing 
$\appcont$. In a rigid tube, however, at the same $\capno$, $\appcont$ continuously decreases until the interfacial instability starts at $t \approx 2$; see the inset in panel (b). As previously observed in a forced drainage case, panel (b) shows that for the same value of $\capno$, $\appcont$ is always smaller in a rigid tube than in a soft tube for the same $\capno$.}

\subsubsection{Effect of the tube's compliance on the air-glycerol interface stability}

\begin{figure}[t] 
    \centering
    \begin{overpic}[width=0.85\linewidth]{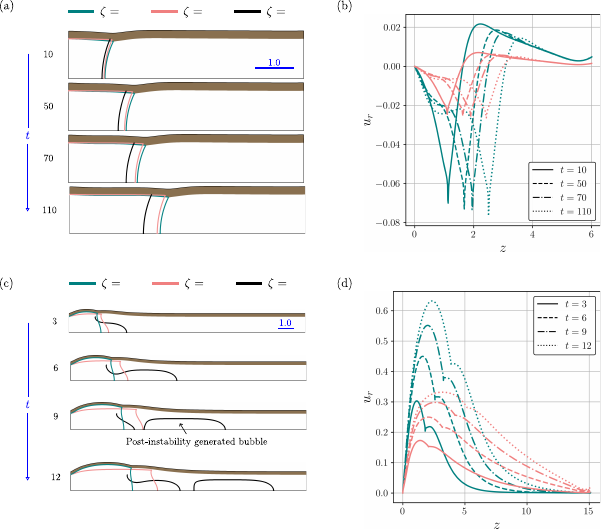}
        \put(20.5,85.6){{\tiny 0.46}}
        \put(34,85.6){{\tiny 1.38}} 
        \put(47.6,85.6){{\tiny $\infty$}} 
        \put(20.8,40.5){{\tiny 0.35}}
        \put(34.5,40.5){{\tiny 2.31}}  
        \put(48,40.5){{\tiny $\infty$}}         
    \end{overpic}
    \caption{{Effect of the tube's compliance on the stability of the air-glycerol interface in a non-wetting capillary tube ($\theta = 115^\circ$). We study two values of $\capno$. Panels (a) and (b) show $\capno=0.01$. Panels (c) and (d) show $\capno=0.5$. (a) Time evolution of air-glycerol interface for different values of $\zeta$. The interface is stable for all values of $\zeta$. (b) Spatio-temporal evolution of the radial solid displacement at the fluid-solid for different values of $\zeta$. (c) Time evolution of air-glycerol interface profiles for different values of $\zeta$. The interface is unstable for $\zeta = 2.31$ and $\infty$, but stable for $\zeta = 0.35$. (d) Spatio-temporal evolution of the radial solid displacement at the fluid-solid interface.}}
    \label{fig:deformability_nonwet}
\end{figure}
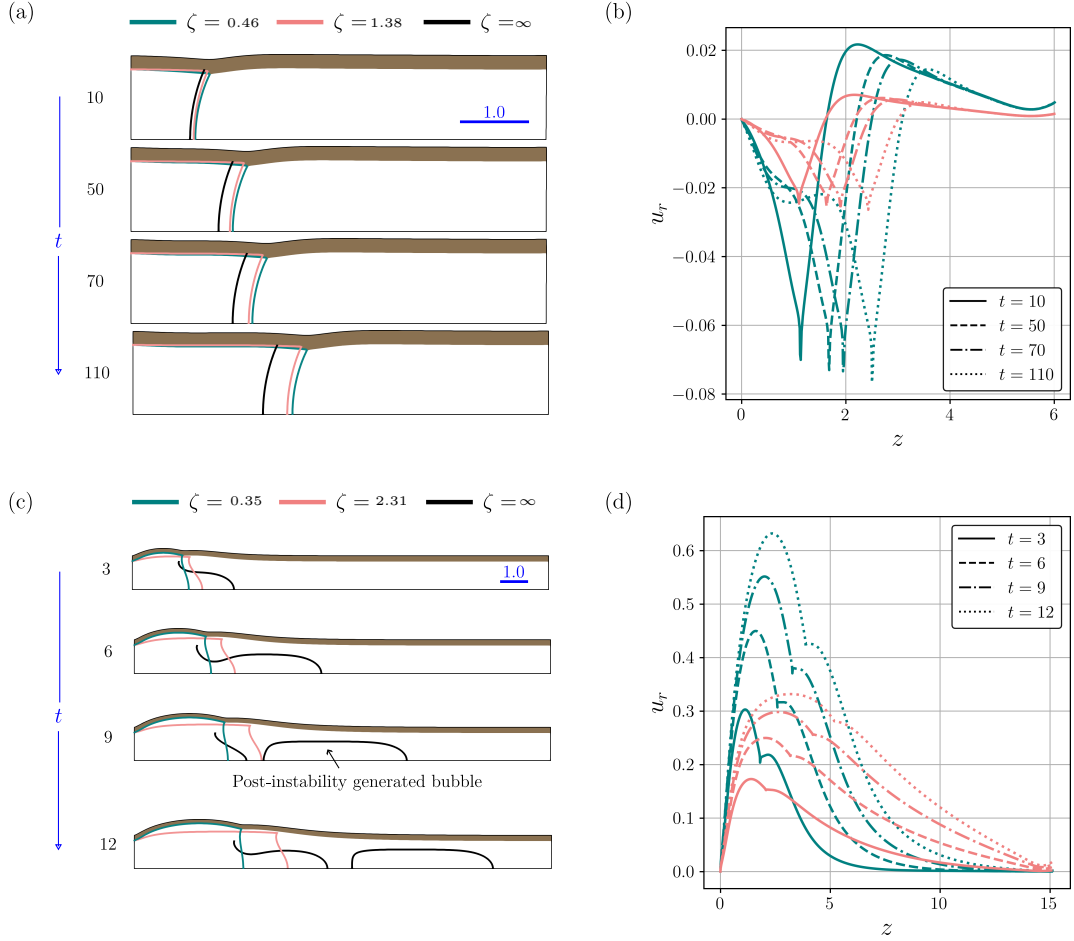

{We investigate the effect of the the tube's compliance on the stability of the air-glycerol interface for two different values of $\capno$: a case when the interface is always stable ($\capno = 0.01$), and a case when the interface is unstable ($\capno = 0.5$); see snapshopts of the solutions in Fig.~\ref{fig:deformability_nonwet}.}

{We first consider $\capno = 0.01$ for three different cases: $\lbr \zeta, \chi\rbr = \lbr 0.46, 1.74 \times 10^{-2}\rbr, \lbr 1.38, 5.80 \times 10^{-3}\rbr$ and $\lbr \infty, 0 \rbr$, the last corresponding to a rigid tube}. For this value of $\capno$, we will use a tube of length $L = 6$. The results are shown in panels (a) and (b). In this system, the interface tip always lags behind the contact line. The  interface velocity increases as $\zeta$ decreases; see panel (a). This occurs because, as $\zeta$ decreases, the tube’s contraction increases, which by mass conservation, leads to a larger local fluid velocity. Panel (b) shows the radial contraction of the tube at the fluid-solid interface, which decreases by a factor of $\approx 3$ as $\zeta$ is increased from $0.46$ to $1.38$. The air-glycerol interface remains stable for all three values of $\zeta$ in this case because the stabilizing effect caused by the air-glycerol interface curvature outweighs any destabilizing effect caused by the tube's contraction.

{We next analyze $\capno = 0.5$ for three different cases: $\lbr \zeta, \chi \rbr = \lbr 0.35, 2.31 \times 10^{-2} \rbr,\, \lbr 2.31, 3.47 \times 10^{-3} \rbr$ and $\lbr \infty, 0 \rbr$. For this value of $\capno$, we will use a tube of length $L = 15$. The results are shown in panels (c) and (d). In this system, the interface tip always leads the contact line. The air-glycerol interface is unstable in the rigid tube and in $\zeta = 2.31$ case, but stable when $\zeta = 0.35$.} The air-glycerol interface advances faster as $\zeta$ increases; see panel (c). The onset of interfacial instability follows the same trend, occurring first in the rigid tube at $t \approx 1$ and then in the $\zeta = 2.31$ case at $t \approx 40$ (data not shown). We also observe that by the time the air-glycerol interface in $\zeta = 0.35$ has advanced one-quarter of the computational domain, one air bubble in the rigid tube has already separated from the injecting air phase. The trend in the air-glycerol interface speed follows the same reasoning as in the forced drainage case---the tube’s expansion, as shown in panel (d), decreases with an increase in $\zeta$, which reduces the local fluid velocity. The reduction in viscous forces caused by the tube’s expansion is not strong enough to overcome the destabilizing influence of the interface curvature when $\zeta = 2.31$ or in the rigid tube; the air-glycerol interface therefore becomes unstable for these two cases. However, for $\zeta = 0.35$, the tube's expansion-induced reduction in viscous forces is sufficient enough to offset the destabilizing influence of the interface curvature, so the interface remains stable.

\subsection{Bubble pinch-off in a soft capillary tube}

\begin{figure}[t] 
    \centering
    \begin{overpic}[width=0.9\linewidth]{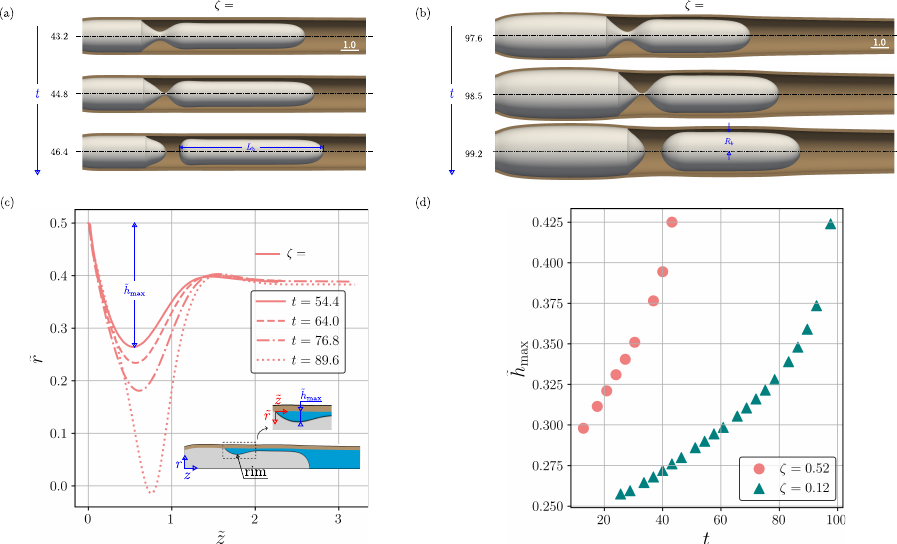}
        \put(79,59.6){{\tiny 0.12}}
        \put(26.5,59.6){{\tiny 0.52}} 
        \put(34.5,32.0){{\tiny 0.52}}
    \end{overpic}
    \caption{Bubble pinch-off in a soft wetting capillary ($\theta = 68^\circ$) tube for inlet capillary number $\capno = 0.1$. Panels (a) and (b) show different stages of the pinch-off process: pre-pinch-off, pinch-off and post-pinch-off, for $\zeta = 0.52$ (a) and $\zeta = 0.12$ (b). The bubble is shown by the level set $\phase = 0$. (c) Time variation of the dewetting rim profile in the comoving frame of the receding contact line, defined along its radial and axial coordinates $\tilde{r}$ and $\tilde{z}$, respectively for $\zeta = 0.52$. (d) Time variation of the maximum height of the rim $\tilde{h}_\text{max}$ for two different values of elastocapillary number $\zeta$. We use $L = 15$ for our computations, but we only display a small section of the tube's length.}
    \label{fig:pinchoff_softtube}
\end{figure}

{In this sub-section, we study the post-instability dynamics in a soft wetting tube at $\capno = 0.1$ for two different cases: $\lbr \zeta, \chi \rbr = \lbr 0.12, 8.67 \times 10^{-3} \rbr$ and $\lbr 0.52, 1.73 \times 10^{-3} \rbr$. For the simulations reported here, we use $R_c = 0.5,\, h = 0.1$ and a tube of length $L = 15$. The air-glycerol interface is unstable in both cases, and we present the corresponding solution snapshots in Fig.~\ref{fig:pinchoff_softtube}.} 

{After the onset of interfacial instability in the $\zeta = 0.12$ and $0.52$ cases, an air bubble pinches-off from the advancing air-phase. Here, we study the effect of $\zeta$ on three pinch-off metrics: i) the timing of bubble pinch-off, ii) the pinched-off bubble length $L_b$ and the iii) average radius of the pinched-off bubble $R_b$. Since the local bubble radius in a soft tube varies in space due to the tube’s dynamic deformation, we have computed $R_b$ by spatially averaging the local bubble radius along the bubble length. Our simulation data in panel (a) shows that the first bubble pinches-off in the $\zeta = 0.52$ case at $t = 44.8$, with $L_b = 5.79$ and $R_b = 0.41$ at $t = 46.4$.  When $\zeta = 0.12$, the bubble pinches-off at $t = 98.4$ as shown in panel (b)---this is much later than in the $\zeta = 0.52$ case. 
As observed with the onset time of interfacial instability, the pinch-off time is inversely correlated to $\zeta$--- this is because bubble pinch-off always occurs after the onset of interfacial instability. When $\zeta = 0.12$, our data show $L_b = 4.14$ and $R_b = 0.58$ at $t = 99.2$, which shows that the lenght and radius of the pinched-off bubble strongly depend on $\zeta$. Because the local capillary number $\capnoavg$ (see \ref{sec:appendix_a} for a definition) increases with $\zeta$ in a soft wetting tube, our observation that $L_b$ reduces as $\zeta$ decreases aligns with the rigid-tube studies showing an increase in $L_b$ with $\capnoavg$ \cite{esmaeilzadeh_etal_pre_2020}. In addition, $R_b$ increases as the tube becomes softer. A softer tube expands more. This makes $\capnoavg$ become smaller, which also leads to 
a smaller $h_f$ (replace $\capnotip$ with $\capnoavg$ in Eq.~\eqref{eqn:filmthickness_exprs_1}). Because the tube's radius is larger and $h_f$ is smaller, the bubble radius grows. A detailed study on the dependence of $R_b$, $L_b$, and pinch-off time on $\capno$, tube wettability, and $\zeta$ is, beyond the scope of the present work, and we leave it to future research.}

{We now explain the mechanism of bubble pinch-off. During an interfacial instability, the contact line behind the entrained film recedes from the tube's wall, forming a dewetting rim; see the inset in panel (c). We plot the axial and time evolution of the rim’s profile for $\zeta = 0.52$ in panel (c); the data for $\zeta = 0.12$ are not shown because they follow the same trend as in $\zeta = 0.52$. To better understand the process of bubble pinch-off, we look at the time evolution of the maximum height of the dewetting rim $\tilde{h}_\text{max}$ in panel (d). At early times for both cases of $\zeta$, $\sfrac{\mathrm{d} \tilde{h}_\text{max}}{\mathrm{d} t}$ is small, implying that the thin film flow near the rim is axially dominated \cite{pahlavan_etal_pnas_2019}. At later times, the flow near the rim becomes radially dominated and $\tilde{h}_\text{max}$ rises rapidly. When $\tilde{h}_\text{max}$ approaches the local tube radius, the air-glycerol interface becomes susceptible to Rayleigh-Plateau instability, subsequently pinching off an air bubble from the advancing air finger.}

\section{Conclusions}

We developed a computational framework to simulate interfacial instabilities and post-instability bubble pinch-off in a soft capillary tube. In our computational approach, we leverage a phase field model for the fluids, a large deformation model for the solid tube, and a body-fitted fluid-structure interaction method. We show that a dynamic wettability condition can quantitatively reproduce the onset of interfacial instability in a rigid capillary tube for imbibition and drainage. To demonstrate the potential of our framework, we study the interfacial instability by varying the inlet capillary number and the elastocapillary number for both imbibition and drainage in a soft capillary tube. The criterion that is normally used to define the onset of the instability in a rigid tube is not valid in a compliant tube. We propose to use the interface arc length and apparent contact angle to quantitatively determine the onset time of the interfacial instability in a compliant tube. Our simulations show that the air-glycerol interface remains stable at low inlet capillary numbers but becomes unstable at higher values. At high values of the inlet capillary number, the instability begins when a finger propagates through the center of the capillary tube, entraining a thin film of glycerol along the tube walls. Following the instability, bubble pinch-off occurs, with the shape and timing of the detached bubbles being significantly influenced by the elastocapillary number. Our simulations also show that interfacial instability and bubble pinch-off are delayed or suppressed in a soft capillary tube compared to a rigid tube. Interestingly, the cause of this delay is not universal. {Imbibition and drainage systems exhibit fundamentally different controlling mechanisms.}

Our work represents a first step towards modeling and better understanding interfacial instabilities in a soft capillary tube, caused by the displacement of a more viscous fluid by a less viscous immiscible fluid. Our simulation results can serve as a basis for future experimental efforts. We believe that future enhancements to our computational framework, such as the use of adaptive refinement procedures \cite{brummelen_ijnme_2021, saurabh_ieee_2023}, will be essential for simulating multi-bubble pinch-off and for accurately resolving small-scale features during the pinch-off phenomenon.

\section*{Author contributions}
\textbf{Sthavishtha R. Bhopalam}: conceptualization (supporting); data curation (lead); formal analysis (lead); investigation (equal); methodology (equal); software (lead); validation (lead); visualization (lead); writing - original draft (equal); writing - review \& editing (equal). \textbf{Ruben Juanes}: conceptualization (supporting); investigation (supporting); supervision (supporting); writing - original draft (supporting); writing - review \& editing (equal). \textbf{Hector Gomez}: conceptualization (lead); formal analysis (supporting); funding acquisition (lead); investigation (equal); methodology (equal); project administration (lead); resources (lead); supervision (lead); writing - original draft (equal); writing - review \& editing (equal). All authors reviewed and approved the final version of the manuscript.

\section*{Declaration of Competing Interest}
The authors declare that they have no known competing financial interests or personal relationships that could have appeared to influence the work reported in this paper.

\section*{Acknowledgements}
SRB acknowledges brief discussions about the computational model with Makrand Khanwale and Tianyi Hu. SRB thanks Yu Qiu for suggesting the use of a dynamic wettability condition. This research was supported by the National Science Foundation (Award no. CBET 2012242). This work used Bridges-2 system at the Pittsburgh Supercomputing Center (PSC) through allocation no. MCH220014 from the Advanced Cyberinfrastructure Coordination Ecosystem: Services $\&$ Support (ACCESS) program \cite{boerner_access_2023}, which is supported by National Science Foundation grant nos. 2138259, 2138286, 2138307, 2137603, and 2138296. The opinions, findings, and conclusions, or recommendations expressed are those of the authors and do not necessarily reflect the views of the National Science Foundation.

\section*{Data availability}
Data for this article are available at \href{https://doi.org/10.4231/6DWP-1274}{https://doi.org/10.4231/6DWP-1274}.

\appendix

\section{Local capillary number in a soft tube}
\label{sec:appendix_a}

\begin{figure*} 
    \centering
    \includegraphics[width=\linewidth]{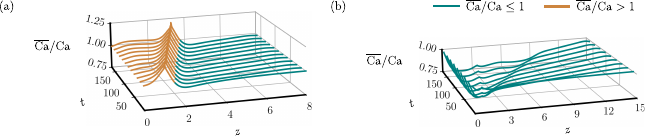}
    \caption{Space and time evolution of the ratio of the average capillary number ($\capnoavg$) to the inlet capillary number ($\capno$) in a soft capillary tube. (a) Forced imbibition case for $\capno = 0.01$, $\theta = 115^\circ$, $\nu = 0.25$ and $\zeta = 0.92$. (b) Forced drainage case for $\capno = 0.04$, $\theta = 68^\circ$, $\nu = 0.45$ and $\zeta = 0.92$.}
    \label{fig:capnovary_softtube}
\end{figure*}

{In the present work, we analyze interface stability primarily with the inlet capillary number $\capno$---the most common definition used in the literature \cite{zhao_etal_prl_2018, pahlavan_etal_pnas_2019, gao_etal_jfm_2019, esmaeilzadeh_etal_pre_2020}. An alternative definition, emanating from mass conservation, is the local capillary number $\capnoavg \lbr z, t \rbr = \capno/\big(R_c' \lbr z, t \rbr\big)^2$. Whereas $\capno$ quantifies the ratio of the viscous forces to the capillary forces at the inlet, $\capnoavg$ quantifies the ratio of the same forces at any axial location of the capillary tube. In a rigid tube, $\capnoavg = \capno$ and $\capnoavg$ is a constant. However, in a soft capillary tube, $\capnoavg$ and $\capno$ are normally different. If the tube expands locally, $R_c' \geq 1$ and $\capnoavg \leq \capno$; and if the tube contracts locally, $R_c' \leq 1$ and $\capnoavg \geq \capno$. Fig.~\ref{fig:capnovary_softtube} shows the evolution of the ratio $\sfrac{\capnoavg}{\capno}$ for both imbibition and drainage cases, where we have used different colors to denote regions of local tube contraction or expansion.}

{Although we do not use $\capnoavg$ in this work, it is an important parameter for three reasons. First, it is an indicator of the local tube deformation. Its ratio with $\capno$ can be an alternative to the radial solid displacement plots computed at the fluid-solid interface. Second, it can be an indicator of the local interface stability. In inclined rigid tubes, for example, the air-glycerol interface can be stable in the initial section of the tube and become unstable downstream or vice-versa \cite{suo_wrr_2024}. Although we do not observe such local transitions in the interface stability in our setup, a critical value of $\capnoavg$---an analogue of $\capnocr$ in a horizontally inclined rigid tube could be used to estimate the onset of local interfacial instability. Third, $\capnoavg$ may be used in the correlations of local film thickness, such as Eqs.~\eqref{eqn:filmthickness_exprs_1} and ~\eqref{eqn:filmthickness_exprs_2}. As the entrainment of glycerol along the walls of the soft tube depends on the balance of local viscous and capillary forces, $\capnoavg$ will be more appropriate in such correlations than $\capno$---this warrants future verification through experiments.}

\bibliographystyle{elsarticle-num} 
\bibliography{ref}

\begin{thebibliography}{10}
\expandafter\ifx\csname url\endcsname\relax
  \def\url#1{\texttt{#1}}\fi
\expandafter\ifx\csname urlprefix\endcsname\relax\def\urlprefix{URL }\fi
\expandafter\ifx\csname href\endcsname\relax
  \def\href#1#2{#2} \def\path#1{#1}\fi

\bibitem{orr_taber_sci_1984}
F.~Orr~Jr, J.~Taber, Use of carbon dioxide in enhanced oil recovery, Science 224~(4649) (1984) 563--569.

\bibitem{yan_etal_aplm_2020}
K.~Yan, J.~Li, L.~Pan, Y.~Shi, Inkjet printing for flexible and wearable electronics, APL Materials 8~(12) (2020) 120705.

\bibitem{whitesides_nat_2006}
G.~M. Whitesides, The origins and the future of microfluidics, nature 442~(7101) (2006) 368--373.

\bibitem{hoffman_jcis_1975}
R.~L. Hoffman, A study of the advancing interface. i. interface shape in liquid—gas systems, Journal of colloid and interface science 50~(2) (1975) 228--241.

\bibitem{bretherton_jfm_1961}
F.~P. Bretherton, The motion of long bubbles in tubes, Journal of Fluid Mechanics 10~(2) (1961) 166--188.

\bibitem{taylor_jfm_1961}
G.~Taylor, Deposition of a viscous fluid on the wall of a tube, Journal of fluid mechanics 10~(2) (1961) 161--165.

\bibitem{huh_jfm_1977}
C.~Huh, S.~Mason, The steady movement of a liquid meniscus in a capillary tube, Journal of fluid mechanics 81~(3) (1977) 401--419.

\bibitem{zhou_prl_1990}
M.-Y. Zhou, P.~Sheng, Dynamics of immiscible-fluid displacement in a capillary tube, Physical Review Letters 64~(8) (1990) 882.

\bibitem{ruiz_etal_jfm_2022}
{\'E}.~Ruiz-Guti{\'e}rrez, S.~Armstrong, S.~L{\'e}v{\^e}que, C.~Michel, I.~Pagonabarraga, G.~G. Wells, A.~Hern{\'a}ndez-Machado, R.~Ledesma-Aguilar, The long cross-over dynamics of capillary imbibition, Journal of Fluid Mechanics 939 (2022) A39.

\bibitem{zhao_etal_prl_2018}
B.~Zhao, A.~Alizadeh~Pahlavan, L.~Cueto-Felgueroso, R.~Juanes, Forced wetting transition and bubble pinch-off in a capillary tube, Physical review letters 120~(8) (2018) 084501.

\bibitem{pahlavan_etal_pnas_2019}
A.~A. Pahlavan, H.~A. Stone, G.~H. McKinley, R.~Juanes, Restoring universality to the pinch-off of a bubble, Proceedings of the National Academy of Sciences 116~(28) (2019) 13780--13784.

\bibitem{esmaeilzadeh_etal_pre_2020}
S.~Esmaeilzadeh, Z.~Qin, A.~Riaz, H.~A. Tchelepi, Wettability and capillary effects: Dynamics of pinch-off in unconstricted straight capillary tubes, Physical Review E 102~(2) (2020) 023109.

\bibitem{gao_etal_jfm_2019}
P.~Gao, A.~Liu, J.~J. Feng, H.~Ding, X.-Y. Lu, Forced dewetting in a capillary tube, Journal of Fluid Mechanics 859 (2019) 308--320.

\bibitem{suo_wrr_2024}
S.~Suo, D.~O’Kiely, M.~Liu, Y.~Gan, Geometry effects on interfacial dynamics of gas-driven drainage in a gradient capillary, Water Resources Research 60~(9) (2024) e2023WR036766.

\bibitem{saffman_prs_1958}
P.~G. Saffman, G.~I. Taylor, The penetration of a fluid into a porous medium or hele-shaw cell containing a more viscous liquid, Proceedings of the Royal Society of London. Series A. Mathematical and Physical Sciences 245~(1242) (1958) 312--329.

\bibitem{spelt_jcp_2005}
P.~D. Spelt, A level-set approach for simulations of flows with multiple moving contact lines with hysteresis, Journal of Computational physics 207~(2) (2005) 389--404.

\bibitem{sui_arfm_2014}
Y.~Sui, H.~Ding, P.~D. Spelt, Numerical simulations of flows with moving contact lines, Annual Review of Fluid Mechanics 46~(1) (2014) 97--119.

\bibitem{zhang_jcp_2020}
J.~Zhang, P.~Yue, A level-set method for moving contact lines with contact angle hysteresis, Journal of Computational Physics 418 (2020) 109636.

\bibitem{jacqmin_jcp_1999}
D.~Jacqmin, Calculation of two-phase navier--stokes flows using phase-field modeling, Journal of computational physics 155~(1) (1999) 96--127.

\bibitem{jacqmin_jfm_2000}
D.~Jacqmin, Contact-line dynamics of a diffuse fluid interface, Journal of fluid mechanics 402 (2000) 57--88.

\bibitem{yue_etal_jfm_2010}
P.~Yue, C.~Zhou, J.~J. Feng, Sharp-interface limit of the cahn--hilliard model for moving contact lines, Journal of Fluid Mechanics 645 (2010) 279--294.

\bibitem{yue_feng_pof_2011}
P.~Yue, J.~J. Feng, {Wall energy relaxation in the Cahn–Hilliard model for moving contact lines}, Physics of Fluids 23~(1) (2011) 012106.
\newblock \href {https://doi.org/10.1063/1.3541806} {\path{doi:10.1063/1.3541806}}.

\bibitem{prokopev_pre_2019}
S.~Prokopev, A.~Vorobev, T.~Lyubimova, Phase-field modeling of an immiscible liquid-liquid displacement in a capillary, Physical Review E 99~(3) (2019) 033113.

\bibitem{anderson_arfm_1998}
D.~M. Anderson, G.~B. McFadden, A.~A. Wheeler, Diffuse-interface methods in fluid mechanics, Annual review of fluid mechanics 30~(1) (1998) 139--165.

\bibitem{gomez_zee_2017}
H.~Gomez, K.~G. van~der Zee, Computational Phase-Field Modeling, Encyclopedia of Computational Mechanics (second Edition), 978-1-119-00379-3, John Wiley \& Sons, Ltd., 2017, pp. 1--35.

\bibitem{abels_etal_2012}
H.~Abels, H.~Garcke, G.~Gr{\"u}n, Thermodynamically consistent, frame indifferent diffuse interface models for incompressible two-phase flows with different densities, Mathematical Models and Methods in Applied Sciences 22~(03) (2012) 1150013.

\bibitem{gomez_etal_frontiers_2023}
H.~Gomez, Y.~Leng, T.~Hu, S.~Mukherjee, V.~Calo, Phase-field modeling for flow simulation, in: Frontiers in Computational Fluid-Structure Interaction and Flow Simulation: Research from Lead Investigators Under Forty-2023, Springer, 2023, pp. 79--117.

\bibitem{khanwale_cpc_2022}
M.~A. Khanwale, K.~Saurabh, M.~Fernando, V.~M. Calo, H.~Sundar, J.~A. Rossmanith, B.~Ganapathysubramanian, A fully-coupled framework for solving cahn-hilliard navier-stokes equations: Second-order, energy-stable numerical methods on adaptive octree based meshes, Computer Physics Communications 280 (2022) 108501.

\bibitem{dong_jcp_2014}
S.~Dong, An outflow boundary condition and algorithm for incompressible two-phase flows with phase field approach, Journal of Computational Physics 266 (2014) 47--73.

\bibitem{dong_shen_jcp_2012}
S.~Dong, J.~Shen, A time-stepping scheme involving constant coefficient matrices for phase-field simulations of two-phase incompressible flows with large density ratios, Journal of Computational Physics 231~(17) (2012) 5788--5804.

\bibitem{simo_2006_book}
J.~C. Simo, T.~J.~R. Hughes, Computational {I}nelasticity, Vol.~7, Springer New York, 1998.

\bibitem{bueno_2018b}
J.~Bueno, H.~Casquero, Y.~Bazilevs, H.~Gomez, Three-dimensional dynamic simulation of elastocapillarity, Meccanica 53~(6) (2018) 1221--1237.

\bibitem{bhopalam_cmame_2022}
S.~R. Bhopalam, J.~Bueno, H.~Gomez, Elasto-capillary fluid--structure interaction with compound droplets, Computer Methods in Applied Mechanics and Engineering 400 (2022) 115507.

\bibitem{donea_2004}
J.~Donea, A.~Huerta, J.-P. Ponthot, A.~Rodr{\'\i}guez-Ferran, Arbitrary Lagrangian--Eulerian Methods, Vol.~3, Encyclopedia of Computational Mechanics, Fluids, 2004, Ch.~14.

\bibitem{wick_2011}
T.~Wick, Fluid-structure interactions using different mesh motion techniques, Computers \& Structures 89~(13-14) (2011) 1456--1467.

\bibitem{bazilevs_fsirev_2008}
Y.~Bazilevs, V.~M. Calo, T.~J.~R. Hughes, Y.~Zhang, Isogeometric fluid-structure interaction: theory, algorithms, and computations, Computational Mechanics 43~(1) (2008) 3--37.

\bibitem{hughes_cmame_1998}
T.~J. Hughes, G.~R. Feij{\'o}o, L.~Mazzei, J.-B. Quincy, The variational multiscale method—a paradigm for computational mechanics, Computer methods in applied mechanics and engineering 166~(1-2) (1998) 3--24.

\bibitem{hughes_etal_2018}
T.~J.~R. Hughes, G.~Scovazzi, L.~P. Franca, Multiscale and stabilized methods, Encyclopedia of Computational Mechanics Second Edition, Wiley Online Library, 2017, pp. 1--64.

\bibitem{bazilevs_2013}
Y.~Bazilevs, K.~Takizawa, T.~E. Tezduyar, Computational fluid-structure interaction: methods and applications, John Wiley \& Sons, 2013.

\bibitem{johnson_book_2009}
C.~Johnson, Numerical solution of partial differential equations by the finite element method, Courier Corporation, 2009.

\bibitem{cottrell_wiley_2009}
J.~A. Cottrell, T.~J. Hughes, Y.~Bazilevs, Isogeometric analysis: toward integration of CAD and FEA, John Wiley \& Sons, 2009.

\bibitem{hughes_cmame_2005}
T.~J. Hughes, J.~A. Cottrell, Y.~Bazilevs, Isogeometric analysis: Cad, finite elements, nurbs, exact geometry and mesh refinement, Computer methods in applied mechanics and engineering 194~(39-41) (2005) 4135--4195.

\bibitem{chung_1993}
J.~Chung, G.~M. Hulbert, {A Time Integration Algorithm for Structural Dynamics With Improved Numerical Dissipation: The Generalized-$\alpha$ Method}, Journal of Applied Mechanics 60~(2) (1993) 371--375.

\bibitem{jansen_2000}
K.~E. Jansen, C.~H. Whiting, G.~M. Hulbert, A generalized-$\alpha$ method for integrating the filtered {N}avier--{S}tokes equations with a stabilized finite element method, Computer methods in applied mechanics and engineering 190~(3-4) (2000) 305--319.

\bibitem{gmres_1986}
Y.~Saad, M.~H. Schultz, Gmres: A generalized minimal residual algorithm for solving nonsymmetric linear systems, SIAM Journal on Scientific and Statistical Computing 7~(3) (1986) 856--869.

\bibitem{tezduyar_cmame_2006}
T.~E. Tezduyar, S.~Sathe, R.~Keedy, K.~Stein, Space--time finite element techniques for computation of fluid--structure interactions, Computer methods in applied mechanics and engineering 195~(17-18) (2006) 2002--2027.

\bibitem{petsc-web-page}
S.~Balay, S.~Abhyankar, M.~F. Adams, J.~Brown, P.~Brune, K.~Buschelman, L.~Dalcin, A.~Dener, V.~Eijkhout, W.~D. Gropp, D.~Karpeyev, D.~Kaushik, M.~G. Knepley, D.~A. May, L.~C. McInnes, R.~T. Mills, T.~Munson, K.~Rupp, P.~Sanan, B.~F. Smith, S.~Zampini, H.~Zhang, H.~Zhang, {PETS}c {W}eb page, \url{https://www.mcs.anl.gov/petsc} (2019).

\bibitem{PETiga_CMAME}
L.~Dalcin, N.~Collier, P.~Vignal, A.~Côrtes, V.~Calo, {PetIGA}: A framework for high-performance isogeometric analysis, Computer Methods in Applied Mechanics and Engineering 308 (2016) 151 -- 181.

\bibitem{style_etal_prl_2013}
R.~W. Style, R.~Boltyanskiy, Y.~Che, J.~S. Wettlaufer, L.~A. Wilen, E.~R. Dufresne, Universal deformation of soft substrates near a contact line and the direct measurement of solid surface stresses, Physical review letters 110~(6) (2013) 066103.

\bibitem{fermigier_jcis_199}
M.~Fermigier, P.~Jenffer, An experimental investigation of the dynamic contact angle in liquid-liquid systems, Journal of colloid and interface science 146~(1) (1991) 226--241.

\bibitem{qiu_thesis_2024}
Y.~Qiu, Fluid-fluid displacement in porous-media microfluidics, Ph.D. thesis, Massachusetts Institute of Technology (2024).

\bibitem{aussillous_pof_2000}
P.~Aussillous, D.~Qu{\'e}r{\'e}, Quick deposition of a fluid on the wall of a tube, Physics of fluids 12~(10) (2000) 2367--2371.

\bibitem{more_report_1980}
J.~J. Mor{\'e}, B.~S. Garbow, K.~E. Hillstrom, User guide for minpack-1, Tech. rep., CM-P00068642 (1980).

\bibitem{sauer_mms_2016}
R.~A. Sauer, A contact theory for surface tension driven systems, Mathematics and Mechanics of Solids 21~(3) (2016) 305--325.

\bibitem{brummelen_ijnme_2021}
E.~H. van Brummelen, T.~H. Demont, G.~J. van Zwieten, An adaptive isogeometric analysis approach to elasto-capillary fluid-solid interaction, International Journal for Numerical Methods in Engineering 122~(19) (2021) 5331--5352.

\bibitem{saurabh_ieee_2023}
K.~Saurabh, M.~Ishii, M.~A. Khanwale, H.~Sundar, B.~Ganapathysubramanian, Scalable adaptive algorithms for next-generation multiphase flow simulations, in: 2023 IEEE International Parallel and Distributed Processing Symposium (IPDPS), IEEE, 2023, pp. 590--601.

\bibitem{boerner_access_2023}
T.~J. Boerner, S.~Deems, T.~R. Furlani, S.~L. Knuth, J.~Towns, Access: Advancing innovation: Nsf’s advanced cyberinfrastructure coordination ecosystem: Services \& support, in: Practice and Experience in Advanced Research Computing 2023: Computing for the Common Good, 2023, pp. 173--176.

\end{thebibliography}





\end{document}